\documentclass[12pt,english]{article}
\pdfoutput=1

\usepackage[authordate,
backend=biber,
doi=only,
isbn=false,
sorting=nyt,
maxcitenames=3,
minbibnames=7,
maxbibnames=7,
uniquename=false,
sortcites=true]{biblatex-chicago}

\bibliography{bib/zotero.bib}

\AtEveryBibitem{\clearlist{note}\clearlist{language}\clearlist{issn}} 
\AtEveryBibitem{%
	\ifentrytype{online}{%
		\clearfield{urlyear}
		\clearfield{urlmonth}
		\clearfield{urlday}
		\clearfield{note}
		\clearlist{language}
	}{%
		\clearfield{eprint}%
		\clearfield{urlyear}
		\clearfield{urlmonth}
		\clearfield{urlday}
		\clearfield{note}
		\clearlist{language}
	}
}%

\usepackage{mathptmx}
\usepackage{microtype}
\everypar{\looseness=-1}
\usepackage{amsmath}
\usepackage{bbm}

\usepackage[utf8]{inputenc}
\usepackage[T1]{fontenc}
\usepackage{babel}

\usepackage{setspace}

\usepackage{breakurl}
\usepackage{graphicx}
\usepackage{booktabs,dcolumn}
\usepackage{array}
\usepackage{multirow}
\usepackage[font=singlespacing, skip=3pt]{caption}
\usepackage[usenames,dvipsnames,svgnames,table]{xcolor}
\usepackage{float}
\usepackage[export]{adjustbox}[2011/08/13]
\usepackage{enumitem}
\usepackage{tikz}
\usepackage{subfig}
\usepackage{float}

\usepackage[hang, flushmargin, bottom, symbol]{footmisc}
\usepackage[colorlinks=true, linkcolor=blue, citecolor=blue, plainpages=false, pdfpagelabels=true, urlcolor=blue]{hyperref}
\usepackage[text={16cm,24cm}]{geometry}
\usepackage{ragged2e}

\usepackage{xspace}

\usepackage[frozencache,cachedir=.,newfloat]{minted}
\SetupFloatingEnvironment{listing}{name=Source Code}
\newcommand{\PARPMTEN}{PM\textsubscript{10}\xspace}

\makeatletter
\date{October 24, 2021}

\newcolumntype{L}[1]{>{\raggedright\let\newline\\\arraybackslash\hspace{0pt}}m{#1}}
\newcolumntype{C}[1]{>{\centering\let\newline\\\arraybackslash\hspace{0pt}}m{#1}}
\newcolumntype{R}[1]{>{\raggedleft\let\newline\\\arraybackslash\hspace{0pt}}m{#1}}

\newcommand{\sym}[1]{\ifmmode^{#1}\else\(^{#1}\)\fi}

\usepackage{calc}
\usepackage{footnotebackref}
\newcommand{\PAPERKEYWORDS}{\textbf{Keywords}: air pollution, birth weight, maternal education, extreme temperatures, China}
\newcommand{\PAPERJEL}{\textbf{JEL}: I14, J1,Q53, Q54}

\newcommand{\PAPERKEYWORDSSEC}{
\section*{Keywords}
\PAPERKEYWORDS
}

\newcommand{\PAPERTITLE}{Same environment, stratified impacts? Air pollution, extreme temperatures, and birth weight in south China}

\newcommand{\AUTHORZHAO}{Qingguo Zhao}
\newcommand{\AUTHORZHAOURL}{https://www.scopus.com/authid/detail.uri?authorId=55766446700}
\newcommand{\AUTHORZHAOINFO}{\href{\AUTHORZHAOURL}{\AUTHORZHAO}: 
Epidemiological Research Office of Key Laboratory of Male Reproduction and Genetics National Health and Family Planning Commission, Family Planning Research Institute of Guangdong Province, Guangzhou, Guangdong, China}

\newcommand{\AUTHORBEHRMAN}{Jere R. Behrman}
\newcommand{\AUTHORBEHRMANURL}{http://orcid.org/0000-0001-7835-6283}
\newcommand{\AUTHORBEHRMANINFO}{\href{\AUTHORBEHRMANURL}{\AUTHORBEHRMAN}: Departments of Economics and Sociology and Population Studies Center, University of Pennsylvania, 133 South 36th Street, Philadelphia, Pennsylvania, USA}

\newcommand{\AUTHORHANNUM}{Emily Hannum}
\newcommand{\AUTHORHANNUMURL}{https://orcid.org/0000-0003-2011-9984}
\newcommand{\AUTHORHANNUMINFO}{\href{\AUTHORHANNUMURL}{\AUTHORHANNUM} (corresponding, \href{mailto:hannumem@soc.upenn.edu}{hannumem@soc.upenn.edu}): Department of Sociology and Population Studies Center, University of Pennsylvania, 3718 Locust Walk, Philadelphia, Pennsylvania, USA}

\newcommand{\AUTHORLIU}{Xiaoying Liu}
\newcommand{\AUTHORLIUURL}{https://orcid.org/0000-0001-9897-1291}
\newcommand{\AUTHORLIUINFO}{\href{\AUTHORLIUURL}{\AUTHORLIU}: Population Studies Center, University of Pennsylvania, 3718 Locust Walk, Philadelphia, Pennsylvania, USA}

\newcommand{\AUTHORWANG}{Fan Wang}
\newcommand{\AUTHORWANGURL}{https://orcid.org/0000-0003-2640-5420}
\newcommand{\AUTHORWANGINFO}{\href{\AUTHORWANGURL}{\AUTHORWANG}: Department of Economics, University of Houston, 3623 Cullen Boulevard, Houston, Texas, USA}

\newcommand{\HIGHLIGHT}{
\section*{Highlights}
\begin{enumerate}
    \item We test maternal education as an effect modifier in associations between air pollution/extreme temperature and birth weight.
    \item We link birth records to environmental data from Guangzhou, China during a period of high and variable air pollution.
    \item Infants with unobserved vulnerabilities---at lower conditional quantiles of birth weight---face more risk from ambient exposures.
    \item The protection associated with college-educated mothers with respect to pollution and extreme heat is substantial.
    \item Protection is amplified under more extreme ambient conditions and for infants with greater unobserved innate vulnerability.
\end{enumerate}
}

\newcommand{\ACKNOWLEDGMENTS}{This paper is part of the project "Prenatal Air Pollution Exposures and Early Childhood Outcomes," which is supported by a grant from the Penn China Research and Engagement Fund (PIs: Behrman and Hannum) and by National Science Foundation Grant 1756738 (PI: Hannum).  The authors also gratefully acknowledge support from the University of Houston Research Fund and, for coverage of Wang's time, Scholar Grant GS040-A-18 from the Chiang Ching-Kuo Foundation.  We thank Yu Xie and James Raymo for the opportunity to receive feedback at the Princeton Research in East Asian Demography and Inequality Seminar. We also thank Guy Grossman, Harsha Thirumurthy, Heather Schofield, and other participants in the Penn Development Research Initiative (PDRI) Seminar for comments. Finally, the authors appreciate the helpful feedback provided by anonymous reviewers.}

\newcommand{\PAPERABSTRACT}{
This paper investigates whether associations between birth weight and prenatal ambient environmental conditions---pollution and extreme temperatures---differ by 1) maternal education; 2) children’s innate health; and 3) interactions between these two. We link birth records from Guangzhou, China, during a period of high pollution, to ambient air pollution (\PARPMTEN and a composite measure) and extreme temperature data. We first use mean regressions to test whether, overall, maternal education is an ``effect modifier'' in the relationships between ambient air pollution, extreme temperature, and birth weight. We then use conditional quantile regressions to  test for effect heterogeneity according to the unobserved innate vulnerability of babies after conditioning on other confounders. Results show that 1) the negative association between ambient exposures and birth weight is twice as large at lower conditional quantiles of birth weights as at the median; 2) the protection associated with college-educated mothers with respect to pollution and extreme heat is heterogeneous and potentially substantial: between 0.02 and 0.34 standard deviations of birth weights, depending on the conditional quantiles; 3) this protection is amplified under more extreme ambient conditions and for infants with greater unobserved innate vulnerabilities.\\
\PAPERJEL}

\newcommand{\PAPERDOIURL}{https://doi.org/10.1016/j.ssresearch.2021.102691}
\newcommand{\PAPERINFO}{
This paper has been accepted for publication at Social Science Research: \url{\PAPERDOIURL}.
}

\begin{document}

\title{
\vspace*{-1.5em}
\onehalfspacing
\PAPERTITLE\thanks{\PAPERINFO}}

\author{
\AUTHORLIU, 
\AUTHORBEHRMAN, 
\AUTHORHANNUM, 
\AUTHORWANG, 
and \AUTHORZHAO\thanks{
\AUTHORLIUINFO;
\AUTHORBEHRMANINFO;
\AUTHORHANNUMINFO;
\AUTHORWANGINFO;
\AUTHORZHAOINFO.
\ACKNOWLEDGMENTS}}

\date{
February 11, 2022}
\maketitle

\vspace*{-1.0em}
\begin{abstract}
\vspace*{-1em}
\singlespacing
\PAPERABSTRACT
\thispagestyle{empty}
\end{abstract}
\clearpage

\doublespacing
\PAPERKEYWORDSSEC
\HIGHLIGHT
\thispagestyle{empty}
\clearpage

\pagenumbering{arabic}
\setcounter{page}{1}
\renewcommand*{\thefootnote}{\arabic{footnote}}

\section{Introduction}

Ambient environmental conditions, including both air pollution and temperature, have been associated with adverse birth outcomes in a number of settings, though findings for temperature remain limited \autocite{shah_air_2011, stieb_ambient_2012, currie_pollution_2013, kloog_using_2015, zhang_temperature_2017, klepac_ambient_2018, melody_maternal_2019, cho_ambient_2020, decicca_when_2020}.  However, some newborns may be more vulnerable than others to the same ambient environmental conditions, due to inequalities in at least two dimensions. First, the pathway from ambient conditions to adverse birth outcomes may vary with maternal education. For example, children in utero with less-educated mothers may be more vulnerable than those of more-educated mothers if their mothers have poorer access to living, work, transportation, and leisure spaces with indoor air filtration and temperature regulation, or if they have limited knowledge of or resources for mitigation strategies.\footnote{Behavioral strategies that respond to extreme pollution and temperature, undertaken by those who can afford them, have been documented in China.  For example, \textcite[518]{zhang_air_2018} estimated that a 100-point increase in the Air Quality Index in China was associated with a  70.6\% increase in expenditures on anti-PM\textsubscript{2.5} masks and a 54.4\% increase for all types of masks. Using evidence from air-purifier sales linked to city data and household-income data, \textcite{ito_willingness_2020} show that higher-income households in China have significantly higher marginal willingness to pay for clean air compared with lower-income households. With regard to temperature, \textcite{yu_temperature_2019} find that while urban households respond to extreme temperatures via increased energy consumption or air conditioner purchases, rural households are unresponsive to temperature fluctuations. The authors attribute this difference in mitigation practices to a lack of resources in rural households.} Second, babies' different innate health vulnerabilities may relate to unequal outcomes in the same environmental context. The relevance of these two dimensions may be particularly great in more extreme environments, and the protective effects of maternal education may be more pronounced for infants with greater unobserved innate vulnerabilities. 

This paper investigates two key questions. First, do children of expectant mothers in
common ambient pollution and temperature environments experience
different birth outcomes, according to their mothers' education? In other words,
are there protective effects of maternal education? Second, are
protective effects of maternal education more pronounced among infants with greater unobserved vulnerabilities---those poorer innate health? We employ mean and conditional quantile regressions to study these questions jointly, with data from a fairly high-pollution environment, for three key dimensions of ambient environmental exposures: air pollution, extreme heat, and extreme cold. 

We focus on birth weight as the outcome of
interest. Birth weight is a consistent correlate of health and other
socio-economic outcomes across the life course \autocite{behrman_returns_2004, black_cradle_2007, behrman_cross-sectional_2015, komlos_pound_2016}. We utilize a database of every 
singleton live birth recorded every day between 2009 and 2011 in one city district with more than one million residents and fairly high air pollution in south
China, for a total of 53,879 birth records. We link the births data
to daily air pollution and meteorological data spanning the entire
prenatal periods for these births. We exploit daily variations in air pollution levels that are caused mainly by random changes in atmospheric conditions and pollutant emissions \autocite{janke_air_2014, he_severe_2019}. The high frequency of measurement of air pollution allows us to control for year-by-month fixed effects, which capture time patterns that might reflect seasonal effects and changes in economic conditions over time that are shared commonly across the city. We calculate daily mean exposure estimates for three monitored air
pollutants, namely particulate matter (\PARPMTEN), sulfur dioxide (SO\textsubscript{2}), and nitrogen dioxide (NO\textsubscript{2}), and also generate a composite index of potential exposure
to all three air pollutants during pregnancy. We also calculate the percentages of days during pregnancy that were extremely hot or cold, to control for deviations in weather from a common seasonal trend that can also have impacts on birth outcomes.

To address our main research questions, we interact a binary variable indicating whether mothers have university education with measures of pollution and extreme temperature exposure to estimate effect
modifications---protective effects---associated with maternal education,
after adjusting for infant sex, a quadratic function of maternal age,
parity, rainfall, and time trends. To our knowledge, this is the first paper to study protective effects jointly in the context of pollution and extreme temperature. Prior research has shown that pollution and extreme temperature might be independently important factors. However, given the potential correlation between them, it is important to analyze protective effects in a joint framework. 

Further, we allow for the protective effects of maternal education to be
heterogeneous across conditional quantiles of birth weight, which map to infants' unobserved innate health vulnerabilities. While the literature on quantile regressions has suggested that birth weight is an ideal variable on which to apply conditional quantile methods \autocite{koenker_quantile_2001}, this analysis is one of the first to adopt conditional quantile methods to analyze the protective effects of maternal education on birth weight.\footnote{Using Chinese data from 16 counties from 2014 to 2018, \textcite{wu_effects_2021} study the effects of PM\textsubscript{2.5} on birth weight in a conditional quantile context. They estimate heterogeneous effects by maternal educational status as well. In contrast to the current paper, their analysis does consider extreme temperature exposures.} Following the literature on
conditional quantile estimation \autocite{abrevaya_effects_2001, arias_individual_2001, abrevaya_effects_2008}, conditional on observed maternal
characteristics, time trends, and environmental exposures, we interpret
newborns to the left of the conditional distribution of birth weight as
having higher levels of unobserved innate health vulnerabilities.\footnote{An alternative to the conditional quantile approach is to use observed variables that measure innate vulnerabilities of children, which has the benefit of clarifying observed variables that might be useful for targeted policies. The benefit of the conditional quantile approach is that it is able to capture the full spectrum of potential sources of vulnerabilities, not just observed characteristics, and does not impose restrictions on the differences in impact magnitudes between less and more vulnerable individuals.} We organize this paper as follows. Section \ref{background-and-research-questions} gives the background and research questions, Section \ref{sec:datamethods} discusses data and methods, Section \ref{results} gives and discusses the results, and Section \ref{summary-and-conclusions} is a discussion and conclusion.

\section{Background and Research Questions\label{background-and-research-questions}}

\subsection{Air Pollution and Birth Outcomes\label{air-pollution-and-birth-outcomes}}

Studies have traced associations between prenatal exposure to various
kinds of air pollutants and adverse birth weight and other birth
outcomes \autocite{shah_air_2011, stieb_ambient_2012, klepac_ambient_2018, melody_maternal_2019}. 
\textcite{currie_traffic_2011} found that the introduction of electronic toll collection greatly reduced both traffic congestion and vehicle emissions
near highway toll plazas and reduced prematurity and low birth weight among mothers within two
kilometers of toll plazas by 10.8\% and 11.8\%,
respectively, relative to mothers two to ten kilometers from  toll plazas. \textcite{hao_air_2016} analyzed birth records between 2002 and 2006 from the
Georgia Department of PUblic Health and found that all traffic-related
pollutants, including NO\textsubscript{2} and PM\textsubscript{2.5}, were associated with preterm birth. 
\textcite{decicca_when_2020} analyzed the impact of a policy that mandated the reduction of power-plant emissions in the eastern United States. The policy greatly reduced emissions of SO\textsubscript{2} and NO\textsubscript{x}, which are the major precursors for PM\textsubscript{2.5}. Their results suggest that the policy reduced the proportion of newborns downwind born prematurely and experiencing low birth weight, and reduced infant mortality for “high-risk” pregnancies associated with observed maternal health conditions such as diabetes, hypertension or eclampsia. Their conceptualization of risk using observed maternal health conditions is related to but distinct from our focus on unobserved innate vulnerabilities of babies themselves.  \textcite{LIU2022101078} exploited exogenous air quality improvement during the Guangzhou Asian Games period to estimate the effects of PM\textsubscript{10}, NO\textsubscript{2}, and SO\textsubscript{2} on birth outcomes. They find that effects vary by pregnancy trimesters, sex, and maternal age.

A large-scale study in Brisbane, Australia, indicated that pregnancy exposures to PM\textsubscript{2.5}, SO\textsubscript{2}, NO\textsubscript{2}, and ozone (O\textsubscript{3}) were associated with increased
risks of low birth weight, as well as pre-term birth \autocite{chen_exposure_2018}. \textcite{chen_incense_2016} find that incense burning is associated with lower birth weight in boys, and their quantile regression also suggests that the negative associations are larger among the lower quantiles of birth outcomes.

Focusing on particulate matter, a recent United States study  analyzed birth certificates data for the period 1999 to 2007 and
found significant birth weight effects associated with gestational
exposures to one form of coarse particulate matter (PM\textsubscript{{10-2.5}}) \autocite{ebisu_exposure_2016}. Similarly, a study of
7,772 mothers in the Netherlands between
2001 and 2005 found that \PARPMTEN exposure during pregnancy was inversely
associated with birth weight \autocite{van_den_hooven_air_2012}. The World Health Organization (WHO) Global
Survey on Maternal and Perinatal Health study examined whether outdoor
PM\textsubscript{2.5} was associated with adverse birth outcomes among 22
countries from 2004 through 2008 \autocite{fleischer_outdoor_2014}. Results showed
that across all countries, PM\textsubscript{2.5} was not associated with preterm birth,
but was associated with low birth weight. In China, the country with the greatest PM\textsubscript{2.5} range among those studied,
preterm birth and low birth weight both were associated with the highest
quartile of PM\textsubscript{2.5} \autocite{fleischer_outdoor_2014}. \PARPMTEN and PM\textsubscript{2.5} concentrations
are highly correlated in China---especially in southern China \autocite{zhou_concentrations_2016}. Studies in Lanzhou, China and Wuxi, China also indicated that
prenatal exposure to \PARPMTEN increases the risk of preterm birth \autocite{zhao_ambient_2015, han_maternal_2018}.

Several recent reviews have synthesized
findings. \textcite{melody_maternal_2019} conducted a systematic review of acute
air quality change studies and found mixed results: there was some
evidence that maternal exposure to acute changes in air quality of
short-to medium-term duration increased the risk of fetal growth
restriction and preterm birth, but the relationship for other adverse
obstetric or neonatal outcomes was less clear. \textcite{shah_air_2011} conducted a broader systematic review of studies of air pollution exposure and birth outcomes.  This review suggested that maternal exposure to SO\textsubscript{2} was
associated with preterm birth, exposure to PM\textsubscript{2.5} was associated with low
birth weight, preterm birth, and small-for-gestational-age birth, and
exposure to \PARPMTEN was associated with small-for-gestational-age birth.
Evidence for NO\textsubscript{2} was inconclusive.  Another systematic review and meta-analysis of 62 studies indicated that
pooled estimates of effects generally suggested associations between carbon monoxide (CO),
NO\textsubscript{2}, and particulate matter and pregnancy outcomes, but also that there
was a high degree of heterogeneity among studies \autocite{stieb_ambient_2012}.\footnote{Early pollution exposures may endure in their impact. \textcite{currie_what_2014} summarized studies from developed and a few developing countries on the short- and long- term effects of in-utero and early-life exposure to pollution. The review highlighted evidence that exposure to pollution significantly impacted a variety of birth, childhood, and later-life outcomes.} 
 
A recent paper systematically reviewed Chinese and English publications
on links between air pollution exposure and adverse pregnancy outcomes
in China \autocite{jacobs_association_2017}. The authors reported that SO\textsubscript{2} was
consistently associated with lower birth weight and more preterm births
and \PARPMTEN was consistently associated with congenital anomalies. Results for NO\textsubscript{2} were
inconsistent, and the authors concluded that further studies were
required on the effects of PM\textsubscript{2.5}, ozone (O\textsubscript{3}), and carbon monoxide (CO).
While a number of the studies adjusted for education or other measures
of socioeconomic status, the review does not report tests of effect
modification by these stratifiers. A more recent study in Guangdong
Province \autocite{liu_association_2019}, which covered the period of 2014 to 2015, reported, among other significant findings, associations with low birth weight for PM\textsubscript{2.5}, PM\textsubscript{10}, SO\textsubscript{2}, NO\textsubscript{2}, and O\textsubscript{3} in the first month and with PM\textsubscript{2.5}, PM\textsubscript{10}, NO\textsubscript{2}, and O\textsubscript{3} in the last month of pregnancy.

\subsection{Extreme Temperature and Birth Outcomes\label{extreme-temperature-and-birth-outcomes}}

Extreme temperature events are
increasing in frequency, duration, and magnitude \autocite{world_health_organization_protecting_2018}. For example, between 2000 and 2016, the number of
people exposed to heat waves increased by around 125 million \autocite{world_health_organization_protecting_2018}. Temperature extremes have been linked to significant reductions in human health, including height and birth weight \autocite{deschenes_climate_2009, deschenes_temperature_2014, ogasawara_early-life_2019}. However, to date, evidence about extreme temperatures and birth outcomes remains limited
\autocite{kloog_using_2015, zhang_temperature_2017, basu_temperature_2018}, and
that which is available is somewhat mixed. For example, a recent analysis of data from
Korea found no significant temperature effect on birth outcomes \autocite{cho_ambient_2020}. A 2017 systematic review  concluded that the evidence
linking preterm birth and low birth weight to ambient temperature was
still very limited and not conclusive, though there were more
examples of adverse estimated effects for high temperatures than for low
temperatures \autocite{zhang_temperature_2017}. A 2020 systematic review covered papers about the associations of air pollution and heat
exposure with preterm birth, low birth weight, and stillbirth in the
United States \autocite{bekkar_association_2020}. Only three papers were identified
that considered heat and birth weight, but all found significant risk.
One, a paper that analyzed 220,572 singleton births for the years 2002
to 2008 from 12 United States sites, found that whole-pregnancy cold or hot
temperature increased term low birth weight risk \autocite{ha_ambient_2017}. A
second paper, conducted among 43,629 full-term but low birth weight
babies and 2,032,601 normal-weight babies in California for the period
1999 to 2013, found that higher long-term apparent temperature exposure
was associated with term low birth weight \autocite{basu_temperature_2018}. The third
reviewed paper, conducted among births in Massachusetts from 2000
through December 2008 \autocite{kloog_using_2015}, showed decreased birth weight
with increased air temperature. 

In China, a study in the city of
Guangzhou found that exposure to either low or high temperatures during
pregnancy was associated with an increased risk of preterm birth \autocite{he_ambient_2016}.  A near-national scale study by \textcite{hu_too_2019} linked extreme heat exposure during pregnancy to significant reductions in multiple dimensions of adult welfare. The authors report that adults who experienced an additional day of high temperature during their prenatal periods, on average, attained 0.02 fewer years of schooling, had a higher risk of illiteracy by 0.18\%, showed a 0.48\% decrease in standardized word-test scores, and were shorter by 0.02 centimeters.

\subsection{Same Ambient Environment, Different Impacts\label{same-ambient-environment-different-impacts}}
There is reason to believe that the effects of ambient air pollution and temperature exposures might vary by maternal education.  First, because more-educated mothers are likely to enjoy better material resources or better knowledge compared to less-educated mothers, more-educated mothers may well adopt mitigation strategies and thus experience lower realized exposures in the same ambient air pollution and extreme temperature environments.  Second, because better-educated mothers, as a group, are likely to have better nutrition and health care access compared to less-educated mothers, they may have more physical capacity to weather the same realized exposures to pollution and extreme temperatures.

There is at least suggestive evidence that the same ambient environment
might be experienced differently for more and less economically
vulnerable groups, such as groups defined by parental
education.\footnote{Positive associations between maternal education and
  birth outcomes, including birth weight but also preterm birth and birth
  length, have been described in a number of studies \autocite{abel_effects_2002, ballon_which_2019}. Studies in the United States have often found significant differences in birth weight for infants borne in more disadvantaged and segregated socio-economic communities with distressed physical environments \autocite{grady_racial_2006, kothari_interplay_2016}.} In a study in Georgia, \textcite{hao_air_2016}
found that associations of traffic-related pollutants, including NO\textsubscript{2} and
PM\textsubscript{2.5}, with preterm birth tended to be stronger for mothers with low
educational attainment. A study in Korea of multiple air
pollutants showed mixed results, but some evidence of socioeconomic disparities in the effects of full-pregnancy exposure at the lowest conditional quantiles of birth weight \autocite{lamichhane_quantile_2020}. \textcite{heo_risk_2019}
conducted a systematic review of papers assessing the effects of particulate
matter on birth outcomes from 2000 to 2019 and found suggestive evidence
that the risk posed was greater for babies of mothers with lower
educational attainment for preterm birth and low birth weight.

However, in a recent review of epidemiological literature about ambient air
pollution and birth weight, Westergaard and her colleagues \parencite*{westergaard_ambient_2017}
identified only a small number of studies that addressed effect
modification by maternal education status. With regard to
temperature effects, \textcite{zhang_temperature_2017} call for more attention to the possibility of individual effect modifiers; maternal education is one important potential stratifier
(see \textcite{basu_temperature_2018, son_impacts_2019} for inconsistent findings on
maternal education as an effect modifier). Generally, the resources
associated with maternal education might be more likely to be brought to
bear under extreme conditions of pollution and temperature, which would
exacerbate the maternal protective effect under these conditions. A
similar logic suggests that the protective effects of maternal education
with respect to pollution and extreme temperature might be most
pronounced for infants with greater unobserved innate vulnerabilities.  

A recent study from China also examines the heterogeneous association between air pollution and birth weight---but does not consider extreme temperatures---using quantile regression and finds that more pronounced effects were observed in lower and intermediate quantiles \autocite{wu_effects_2021}. This paper observed stronger associations among well-educated mothers. Although the paper uses data from 16 counties across China during the period of 2014 to 2018, it does not control for county-specific time trends that may capture local policy variations and trends in the economic environment that can also affect birth outcomes. Additionally, \textcite{wu_effects_2021} consider only full-term births. Except for \textcite{wu_effects_2021}, we have not found other studies of pollution and birth weight that investigate variability in the protective effect of maternal education using a conditional quantile approach. 

A handful of studies in the United States have used conditional quantile approaches to consider heterogeneous vulnerability to ambient pollution exposure. A study of PM\textsubscript{2.5} in California \autocite{schwarz_quantile_2019} shows a tendency of somewhat larger negative associations with outcomes at the lowest conditional quantiles, though primarily among the non-Hispanic Black population.  A study of PM\textsubscript{2.5} in Massachusetts \autocite{fong_fine_2019} found that the negative association between
PM\textsubscript{2.5} and birth weight was larger in magnitude at the
lower conditional quantiles of birth weight than in the
higher quantiles. In contrast, a study in Atlanta \autocite{strickland_associations_2019} showed larger effects for several pollutants at higher conditional quantiles.  While results are mixed, all of these studies suggest the potential value in testing for heterogeneity in associations across the distribution of children's unobserved innate vulnerabilities at birth, along the conditional distribution of birth weight. 

\subsection{Research Questions}

Using a case study of three years of singleton live births in one district in
Guangzhou, China, we study the protective effect of mothers' education
with respect to pollution and extreme temperature exposure. Specifically, we consider the
association between expectant mothers' college educational attainment
and their children's birth weight and allow the college educational gradient to be
heterogeneous at different ambient pollution and extreme temperature
levels. By including pollution and extreme temperature in the same specifications, we are able to isolate the separate relationships between these correlated but distinct ambient environmental factors and birth outcomes. Conditional on environmental exposures, we also allow the
educational gradient to be heterogeneous for children with different
unobserved innate vulnerabilities.

Our analysis relaxes the restrictions imposed by common regression
frameworks on how socioeconomic factors might impact child-birth
outcomes in a setting where mothers are exposed to high levels of
negative environmental factors. A common regression
framework includes educational status as an additive variable (for
example, see \textcite{koenker_quantile_2001} (birth weight) and \textcite{hanandita_double_2015} (BMI) in the context of quantile
regressions; see \textcite{bharadwaj_atmospheric_2008} in the context of mean
regressions). In the context of (conditional) mean regressions, this
framework means that the educational gradient in birth weight is
constant across levels of environmental exposure levels. By interacting
mothers' education with multiple potential negative environmental
factors, we allow for the educational gradient in child birth outcomes
to be magnified at different rates depending on how negative
environmental factors jointly worsen. Furthermore, by estimating
education-environment interactions separately at different conditional
quantiles of birth weight, we allow the educational gradient to have
different slopes by quantiles of unobserved innate
vulnerabilities of babies. The rich heterogeneous relationships allow
for a nuanced look at the potential protective effects of maternal college education in ameliorating negative environmental effects.

\section{Data and Methods\label{sec:datamethods}}

\subsection{Study Site}

The study site for this project is Guangzhou, which is the capital city
of Guangdong Province in south China, in the Pearl River Delta.
Guangzhou, along with several other cities in Guangdong as well as Hong
Kong and Macau, is part of the rapidly-developing ``Greater Bay Area''
megalopolis. The region is wealthy, with Shenzhen and Guangzhou,
respectively, ranked third and fourth highest in city GDP in China,
after Beijing and Shanghai \autocite{buchholz_infographic_2019}. Guangzhou has a typical
subtropical climate, with very mild winters and hot, humid summers. The
annual mean temperature in Guangzhou is around 22 degrees Celsius
\autocite{climate-dataorg_guangzhou_2020}. In recent decades, Guangzhou has experienced increases in mean annual temperatures and in the frequency, duration, and intensity of heat waves \autocite{zhang_trends_2017}. Extreme heat events are a particularly significant threat to human health due to a combination of the subtropical climate and urban heat-island effects exacerbated by rapid population growth in recent years \autocite{zhang_trends_2017}. Recent studies in Guangzhou have linked maternal heat exposures to preterm births \autocite{he_ambient_2016, wang_independent_2020}. At the same time, and perhaps somewhat counter-intuitively in a subtropical climate, cold spells in Guangzhou carry health risks.  For example, cold spells have  been linked to increased mortality risk for most categories of deaths and to heightened risk of preterm birth \autocite{Chen_2021,he_ambient_2016}, possibly due to the lack of widespread access to central heating. As a major city in south China, Guangzhou is a data point that is potentially relevant to other urban settings in south China, and possibly parts of Southeast Asia, characterized by broadly similar climates, high levels of pollution, and economic conditions.

South China has had less air pollution than north China, but as an economic development center, Guangzhou has been one of the most polluted cities in the region.  Recent studies in Guangzhou have linked air pollution to respiratory distress syndrome and to student respiratory illness and absenteeism, especially for younger students \autocite{chen_air_2018, lin_ambient_2018}.
Importantly, Guangzhou has a well-established and extensive air-quality
monitoring system with both air pollution indices and individual
pollutant concentrations available at the city level.

Figure \ref{fig:mainone} shows levels of air pollution in Guangzhou from 2008 to 2011
for three monitored pollutants: \PARPMTEN, SO\textsubscript{2}, and NO\textsubscript{2}. The levels of pollution depicted in Figure \ref{fig:mainone} reflect
substantial improvements in Guangzhou's air quality compared to prior years. In the early 2000s, the local government implemented progressive air-quality
control efforts that included closing down low-efficiency coal-power plants and enforced installation
of desulfurization facilities \autocite{zhong_sciencepolicy_2013}. Even with improvements, annual mean air pollution in Guangzhou during this period exceeded the WHO standard for PM\textsubscript{10}: a 20 $\mu g/m^3$ annual mean.  Further, 66.4\% of days surpassed the 24-hour mean standard for PM\textsubscript{10} of 50 $\mu g/m^3$ and 76\% of days surpassed the SO\textsubscript{2} standard of 20 $\mu g/m^3$   \autocite{world_health_organization_ambient_2018}. There is strong
seasonality in all three types of air pollution, with more severe air
pollution in the winter than in the summer.

\subsection{Data\label{data}}

We link three forms of data for this analysis: birth certificate data,
air pollution data, and meteorological data.

\subsubsection{Birth Certificate Data \label{birth-certificate-data}}

We have access to birth certificate data representing all births that took place in one district in
the center city of Guangzhou on every day in the years 2009 to 2011. It is the responsibility of the parents or the family concerned
to register all births within 15 days after birth \autocite{lin_association_2015}.  We elected to use birth certificate data from one particular district in the city because of the availability only for that district of information on maternal education.
In addition to maternal education, the birth certificate data include information about gestational age based
on reported last menstrual period and confirmed by ultrasound scanning,
stillbirth, birth weight, birth length, sex, parity, Apgar scores, and
mother's age, occupation, rural/urban residence status, and number of
pregnancies.\footnote{Birth certificates also contain information on
  infant health and neonatal mortality.}  
\subsubsection{Air Pollution Data \label{air-pollution-data}}

Our interest lies in tracing different experiences in a common ambient environment, and we operationalize this idea with a city average pollution measure.  Our data source for air pollution is the Guangzhou Environmental Bureau,
which reports the daily average levels of three monitored ``criteria''
air pollutants, \PARPMTEN, NO\textsubscript{2}, and SO\textsubscript{2}, at the city level during the period
2005 to 2011. These pollutants are measured according to the National
Standard (\emph{Guo Biao)} GB3095---1996. These are the only three pollutants for which data were collected until
2014.
These are measures of ambient conditions, rather than individual pollution exposure measures.\footnote{Pregnant women may come from any district in the city.  There is precedent
  for using city-average measures of pollution. \textcite{zhao_ambient_2015}
  report seven studies that used city-level averages of \PARPMTEN \autocite{sagiv_sharon_k_time_2005, hansen_maternal_2006, jiang_time_2007, darrow_ambient_2009, suh_different_2009, zhao_effects_2011, schifano_effect_2013-1}.}  For our research question, we want ambient measures because women with different levels of education may be able to alter their individual exposure so that individual exposure data might miss some of the impacts of maternal education. 
 
\subsubsection{Meteorological Data\label{meteorological-data}}

We used the universal thermal climate indices (UTCI) from Copernicus and the European Centre for Medium-Range Weather Forecasts (ECMWF) to measure
extreme temperatures \autocite{copernicus_universal_2020}. The UTCI, a thermal comfort indicator
based on human heat-balance models, is designed to be applicable in all
seasons and climates and for all spatial and temporal scales \autocite{copernicus_universal_2020}.
The UTCI is a one-dimensional index that reflects "the human
physiological reaction to the multidimensionally defined actual outdoor
thermal environment" \autocite[481]{brode_deriving_2012}. Scores can be classified
into ten thermal stress categories, ranging from extreme cold stress to
extreme heat stress \autocite{copernicus_universal_2020}. In a subtropical, humid environment
such as Guangzhou, this index provides a wider range of variation than
ambient temperature. These data have high spatial and temporal resolution, with measurement gridded at 0.25° x 0.25° and at hourly frequency. We obtain the measurement at the grid closest to the center city of Guangzhou and calculate the daily average from 2008 to 2011. 

Finally, we use 24-hour accumulated rainfall from the same grid for the same period. We control for rainfall because it may be correlated with pollution and temperature and may have effects on birth outcomes.\footnote{We provide details on how we acquire and process ECMWF data in Appendix \ref{sec:appdata}.}
Accumulated rainfall is calculated from observed rainfall
data, and comes from the  \textcite{national_oceanic_and_atmospheric_administration_global_2020}, as distributed by \textcite{raspisaniye_pogodi_ltd_weather_2020}.

\subsection{Analytic Sample\label{analytic-sample}}

We begin with birth-certificate data for all live births (67,108 or
99.1\% of all births) in the district for the period 01 January 2009 to
31 December 2011. To avoid fixed-cohort bias,\footnote{Fixed-cohort bias
  emerges when a sample consists of births during a fixed period---this
  approach will include only the longer pregnancies at the start of the
  study and only the shorter pregnancies at the end of the study. This
  has the potential to bias studies of environmental exposures \autocite{strand_methodological_2011}.} we delete
births within this period with conception dates earlier than 14 July
2008 and those with conception dates later than 15 February 2011, as
gestational age varies between 171 and 319 days, which leaves 58,827
observations. We exclude observations with birth weight less than 500 grams (17 observations dropped), or birth length less than 28 centimeters or longer than 60 centimeters (51 observations dropped). We further restrict our sample to
include only those with observed gestational age (855 observations
without gestational age dropped) and only live singleton
births.\footnote{Because there is no information on the number of
  multiple births, we count duplicate observations in terms of all
  observed maternal and paternal characteristics (including mother's
  birthdate, parity, number of previous pregnancies, maternal occupation, maternal education, paternal
  education, gestational age, and an indicator of high risk levels) as well as
  birthdate of the baby, and treat the observations with one duplicated
  case as twins (3.90\% of 57,108 observations) and those with two
  duplicated cases as triplets (0.08\% of 57,108 observations) and those
  with three duplicated cases as quadruplets (0.01\% of 57,108
  observations).} We further removed those registered with rural residence (949 observations, or 1.73\% of the sample). In the end, our analytic sample consists of 53,879 live
singleton births (15,068 born in 2009, 20,368 born in 2010, and 18,443
born in 2011).\footnote{In Appendix Section \ref{sec:appfullterm}, we estimate the model after restricting the analytic sample to only those with more than 36 weeks of gestational age.}

\subsection{Measurement\label{measurement}}

Table \ref{tab:mainone} contains summary statistics for all variables employed in the
analysis. Below, we describe our variables.

\subsubsection{Birth Weight and Other Birth Outcomes\label{birth-weight}}

Our study focuses on birth weight.\footnote{Our analysis in the main text of the paper focuses on conditional mean and conditional quantile regressions for birth weight. In Appendix Section \ref{sec:appbinary}, we present results based on the binary outcome variables. For the binary outcomes, low birth weight is 1 if birth weight is under 2500 grams, preterm equals to 1 if gestational age is smaller than 37 weeks, and small-for-gestational-age is defined as 1 for those with birth weight under 10\% in sex- and gestational-age- specific distribution. We use the cutoffs listed in \textcite{Zhu_chinese_2015} as references.} Birth weight is reported in grams. The mean birth weight is 3,182 grams
and the standard deviation is 473 grams. Babies of mothers with a college degree have, on
average, a 46 gram greater birth weight than babies of mothers with at
most a high school degree  (p=0.00). 

\subsubsection{Air Pollution\label{air-pollution}}

We obtained data on three criteria pollutants, \PARPMTEN,
\emph{NO\textsubscript{2}}, and
\emph{SO\textsubscript{2}}, observed at all monitoring stations in Guangzhou 
from the Guangzhou Environmental Bureau. A small number of observations with missing data
are replaced with moving averages for the four preceding and four following days. We calculate the average level of all monitoring stations from the four contiguous districts located in the center of the city, one of which is where our birth data is sourced.\footnote{Table \ref{tab:mainone} provides \PARPMTEN summary statistics from all city districts as well as from the four center-city districts. We use the center-city districts data for estimation. In Appendix Section \ref{sec:appallcity}, we present robustness checks where we use all city districts pollution data.} Between the
years 2009 and 2011, by our estimates, central Guangzhou's air pollution far
exceeded the safety standards set by the WHO, with annual mean
concentrations of \PARPMTEN of 75.9 $\mu g/m^3$ in  2008, 74.4
$\mu g/m^3$ in 2009, 71.2 $\mu g/m^3$ in 2010
and 69.6 $\mu g/m^3$ in 2011. The WHO
safety standard for \PARPMTEN is that the annual mean should not exceed 20 $\mu g/m^3$ \autocite{world_health_organization_ambient_2018}. Seasonal variation is
especially significant. For example, the daily mean for \PARPMTEN is 49.2
$\mu g/m^3$ in July, on average, but 103.3
$\mu g/m^3$ in December, during the period covered by this study.

We exploit the temporal variation in daily air pollution  and the variation in pregnancy timing for identification.  We summarize over the duration of pregnancy
the daily average level of each pollutant to calculate the
accumulated total potential exposure during pregnancy for each individual. However, this
approach suffers from a potential problem---it generates a spurious
inverse correlation between adverse birth outcomes and accumulated air
pollution levels because a shorter gestational age implies a shorter
pregnancy and therefore smaller accumulated air pollution exposure. We
avoid this spurious relationship by dividing the accumulated levels of
each pollutant during pregnancy by each woman's pregnancy duration to
get a daily mean level of exposure during pregnancy.

Because pollutants tend to co-vary, we conduct principal component
analysis of the three pollutants' daily mean potential exposures during
pregnancy and adopt the first principal component, which accounts for
86\% of the total variance, as a composite index of pollution.\footnote{We adopt principal component analysis as a robustness check to investigate whether a more comprehensive air pollution index shows a consistent pattern of contribution to adverse birth outcomes. However, we are also aware that results from the principal component analysis may not be extrapolated to other places as different cities may have different compositions and interactions of air pollutants.} We focus on
results for our particulate matter measure---\PARPMTEN---in the main text,
and comment briefly on similar findings using the composite pollution
index. Detailed results using the composite measure are included in an appendix.

Substantial individual variation in potential pollution exposure comes from
variation in pregnancy timing. As pollution is usually higher in winter
than in summer, those whose pregnancies occur mainly in the winter have
higher ambient pollution exposure than those whose pregnancies occur mainly in
the summer. Moreover, pregnancies in earlier years on average have
higher ambient pollution exposure than those in later years, as a
result of air-quality improvement over time. In the Guangzhou data, there is a possibly counterintuitive pattern in which mothers with at least a college degree have higher potential air
pollution exposure during their pregnancies than mothers with at most a high school degree (as
shown in Table \ref{tab:mainone}). This pattern emerges because mothers with at most a high school degree, a category that includes many migrant workers from rural areas, are more likely to marry and conceive in January and February, around Chinese New Year.  Figure \ref{fig:maintwo} shows that, after controlling for conception
year-by-month fixed effects, the distributions of ambient air pollution exposure
are not significantly different between the two maternal-education
groups.\footnote{Individuals of different socioeconomic backgrounds may differ in their capacities to choose where to live and work based on air quality and to adopt mitigation strategies such as wearing masks or installing air filters. These are possible pathways connecting a common ambient environment to different impacts.  Available data do not permit measurement of realized individual-level air pollution exposures.}

\subsubsection{Extreme Temperature\label{extreme-temperature}}

We set a cutoff for defining extreme temperatures by creating a local
reference dataset comprised of the nine years of records of daily mean
UTCI that immediately preceded the dates of our study. We define extreme
low and high temperatures, respectively, by generating cut-points
marking the bottom 1\% of the distribution of the reference dataset, at
-0.87 degrees Celsius, and the top 1\% of the reference dataset,
at 34.23 degrees Celsius. Each woman's exposure to extreme temperatures
is defined by the proportion of days during pregnancy with daily mean
UTCI under or above the cut-off temperatures. Extreme cold and hot
temperatures defined in this way correspond to ECMWF categories of
moderate cold stress (defined as -13 to 0 degrees Celsius) and strong heat stress (defined as 32 to 38 degrees Celsius). Although air
conditioning had become widespread in Guangzhou by the time of the study
period, heating in winter was not widely available, which is important
for understanding any observed effects of cold stress in this region. Using our definitions, pregnant women in the sample, on average, spent 1.70\% of
their pregnancy duration (4.6 days) exposed to extreme cold and 1.73\%
of the duration of their pregnancy (4.7 days) exposed to extreme heat. 

We perform sensitivity analyses with a less stringent
definition of extreme temperatures, using a 2.5\% cutoff to define
extreme cold and extreme heat (at 2.52 degrees Celsius and 33.56 degrees Celsius,
respectively). Using this alternative definition, on average, pregnant women in the sample spent 4.40\% (11.9 days) of their pregnancy duration in extreme cold and 3.73\% of their pregnancy duration
(10.2 days) in extreme heat. Table \ref{tab:maintwo} presents
additional distributional details of the extreme heat and cold measures. Figure \ref{fig:mainonetemp} presents the time-series of daily mean UTCI between the years 2008 and 2011. 

\subsubsection{Maternal Education\label{maternal-education}}

We code \emph{mothers' education} in two categories: high school and
below (0) and college and above (1). In our analytic sample, 17,967
(33\%) report college attainment. This sample of mothers from the center of the most developed metropolitan city 
in south China is highly educated. Recent census figures for the population ages six and above in Guangdong Province indicate a 16\% rate of tertiary attainment; rates are higher for China's wealthiest province-level cities (Beijing, 42\%, and Shanghai, 34\%).\footnote{Calculated from \textcite{national_bureau_of_statistics_of_china_communique_2021}.}

\subsubsection{Control Variables\label{control-variables}}

Pregnancy risks associated with advanced maternal age are well
established for mothers and their children, and include heightened risk of
pre-term labor, fetal growth restriction, and fetal demise among those
over 35, in comparison to younger mothers \autocite{sauer_reproduction_2015}.\footnote{We have
  found few studies that identify whether advanced maternal age places
  mothers at particular risk to conditions of air pollution exposure.
  However, one cohort study in Wuxi, China showed an effect modification
  with maternal age: preterm birth risk associated with exposure to high
  levels of \PARPMTEN occurred primarily among women over age 35 \autocite{han_maternal_2018}.} Pregnancy when young also tends to be associated with risk, but
often for social as well as biological reasons. To capture non-linear
age effects, we include continuous controls for \emph{maternal age} and the square of
\emph{maternal age}. We adjust for \emph{sex of the child} (0 if
male, 1 if female) and \emph{parity} (with a set of binary dummy
variables). We also include controls for \emph{conception year by} \emph{month}
fixed effects to control for seasonal effects and any time trend\footnote{Controlling for seasonal trends is a common specification in studying the health impacts of ambient environment. Depending on the structure of the data, the literature uses different ways of controlling, varying from calendar day in a year trends \autocite{chen_effects_2020}, to year-by-month fixed effects to exploit within-calendar month variation in the ambient environment \autocite{janke_air_2014}, to year and month fixed effects to exploit cross-year variation in ambient environment in the same month \autocite{currie_weathering_2013}. To exploit the daily variation in air pollution level, we choose to include year-by-month fixed effects to control for seasonal effects and changes in economic conditions over time that are shared commonly across the city that may also confound the results.}, as
well as \emph{day of the week at birth} fixed effects \autocite{he_ambient_2016}. Finally, we
adjust for \emph{daily mean rainfall} during the duration of pregnancy
using a cubic function of daily mean rainfall in all regressions to
model the nonlinear relationship between rainfall and birth weight.\footnote{Rainfall has been related to birth outcomes in a number of studies \autocite{currie_weathering_2013, rocha_water_2015}. Given this precedent and the monsoon climate in Guangzhou, which is often exposed to typhoons, we control for rainfall in our study. Relative humidity is also a common confounder for birth outcomes \autocite{he_ambient_2016, rich_differences_2015}. However, since our temperature variable, UTCI, by construction, has already incorporated both dry bulb temperature and relative humidity, we do not include relative humidity as a separate control in our analysis. Nevertheless, our results are robust if we drop rainfall and include relative humidity in regressions.}

\subsection{Analytic Approach\label{analytic-approach}}

We model birth weight as related to maternal education, air pollution and extreme
temperature exposure in both mean and quantile regression frameworks. We
first estimate a baseline main effects (conditional) mean regression model:
\begin{align}
\label{eq:one}
\begin{split}
\text{BirthWeight}_{\text{ymi}}\thinspace=&\thinspace\thinspace\thinspace 
\alpha + \zeta_{ym} + \sigma_{0}\text{Edu}_{i}\\
&\thinspace 
+ \gamma_{1}P_{i} + \gamma_{2}\text{Cold}_{i} + \gamma_{3}\text{Heat}_{i}\\
&\thinspace
+ g\left( \text{Rainfall}_{i} \right) + \mathbf{X}_{i}\mathbf{\theta}^{'} + \epsilon_{i}
\thinspace,\\
\end{split}
\end{align}
where \(\text{BirthWeight}_{\text{ymi}}\) is birth weight of individual
\(i\) who is conceived in year \(y\) and month \(m\). \(\text{Edu}_{i}\)
is a binary variable indicating if a mother has at least a college degree.
\(P_{i}\) is a continuous variable measuring each mother's prenatal
ambient exposure to PM\textsubscript{10}, or the composite index of
PM\textsubscript{10}, NO\textsubscript{2} and SO\textsubscript{2}.
\(\text{Cold}_{i}\), and \(\text{Heat}_{i}\) are continuous variables
measuring the percentages of prenatal days exposed to extreme cold or
heat. \(g\left( \text{Rainfall}_{i} \right)\) is a cubic function of
daily mean precipitation that a mother is exposed to during her
pregnancy. We control for an individual-specific vector of
attributes \(\boldsymbol{X}_{i}\), which includes linear and quadratic terms for
maternal age, child's sex, parity,  number of
multiple births, and birth day of the week. To control for shifts in
seasonal patterns of birth as well as possible changes in economic conditions over time, we include
conception year and month interaction fixed effects
\(\zeta_{ym}\).\footnote{Due to significant day-to-day variations in pollution exposures as shown in Figure \ref{fig:mainone} and Figure \ref{fig:maintwo}, despite the inclusion of year and month interaction fixed effects, there remains sufficient variations in pollution exposures that allow for identification. These variations, in effect, arise due to day-by-day differences in exposures in the first and last months of pregnancy. In our regression framework, we assume homogeneous effects of pollution exposures during the course of pregnancy.} Given these, Equation \eqref{eq:one} compares the birth
weights of infants conceived in the same calendar month and examines the
associations of cumulative within-month variation in air pollution and
temperature with variation of birth weights.

Equation \eqref{eq:one} restricts the birth weight gradient with  college education 
to be constant, which is captured by \(\sigma_{0}\) and is invariant of ambient
environmental exposures. Equation \eqref{eq:two} relaxes this restriction, and allows the
 birth-weight gradient with college education to potentially vary
depending on \(P_{i}\), \(\text{Cold}_{i}\), and \(\text{Heat}_{i}\):
\begin{align}
\label{eq:two}
\begin{split}
\text{birth weight}_{\text{ymi}}\thinspace=&\thinspace\thinspace\thinspace 
\alpha + \zeta_{ym} + \sigma_{0}\text{Edu}_{i} \\
&\thinspace 
+ \gamma_{1}P_{i} + \gamma_{2}\text{Cold}_{i} + \gamma_{3}\text{Heat}_{i}\\
&\thinspace 
+ \sigma_{1}\text{Edu}_{i} \times P_{i} + \sigma_{2}\text{Edu}_{i} \times \text{Cold}_{i} + \sigma_{3}\text{Edu}_{i} \times \text{Heat}_{i}\\
&\thinspace
+ g\left( \text{Rainfall}_{i} \right) + \mathbf{X}_{i}\theta^{'} + \epsilon_{i}
\thinspace,\\
\end{split}
\end{align}
where positive values for \(\sigma_{1},\ \sigma_{2}\text{\ \ }\)and
\(\sigma_{3}\) combined with negative values for
\(\gamma_{1},\ \gamma_{2}\text{\ \ }\)and \(\gamma_{3}\) would imply
protective effects of maternal college education that ameliorate the
negative impacts of ambient air pollution, extreme cold, and extreme heat
on birth weight.

We estimate Equations \eqref{eq:one} and \eqref{eq:two} both via conditional mean (OLS) and conditional quantile estimation.\footnote{Under OLS, the same
  estimated coefficients provide both conditional and unconditional mean
  predictions; under quantile estimations, conditional quantile
  estimates in general can differ from unconditional quantile results
  \autocite{firpo_unconditional_2009}. In this paper, we use conditional
  quantile estimates to study heterogeneous maternal education and
  birth weight gradients on levels (grams) of birth weight. While \textcite{firpo_unconditional_2009} show that in the
  context of birth weight, conditional and unconditional quantile
  estimates can be similar, we do not consider the implications of our
  estimates for the unconditional (marginal) distribution of
  birth weight. We estimate quantile regressions with the quantreg package \autocite{koenker_quantreg_2020}.} Under OLS, Equation \eqref{eq:two} imposes the restriction that the
birth weight-college education gradient---conditional on observed variables 
including the vector of environmental exposure measures \(P_{i}\),
\(\text{Cold}_{i}\), and \(\text{Heat}_{i}\)---is constant across
individuals. The implicit assumption is that the birth weight-college education gradient is invariant across the conditional distribution of birth weight, where the conditional distribution captures unobserved
factors---including nutrition, avoidance behavior, genetic factors, or
other risk factors---that can also impact birth weight. However, given the same ambient environment, the tendency for
college-educated mothers to have babies with higher birth weight compared
to non-college-educated mothers may be most pronounced at the lower
conditional quantiles of unobserved innate vulnerabilities.
Additionally, babies with greater unobserved innate
vulnerabilities might be more likely to suffer from air pollution and
extreme temperatures.

Our conditional quantile analysis allows  birth
weight-college education gradients to vary both along observed dimensions of sources of
vulnerabilities, i.e., potential negative environmental factors (as seen in the interaction effects in Equation \eqref{eq:two}), and along
unobserved dimensions of sources of baby vulnerabilities, as captured by
different conditional quantiles. The results provide a finer
disentangling of heterogeneities of the birth
weight-maternal education gradients. For example, the conditional mean estimates could
understate the importance of \PARPMTEN if the negative relationship between
\PARPMTEN and birth weight is magnified at lower conditional quantiles where
mothers/children have greater unobserved innate vulnerabilities.

\section{Results\label{results}}

\subsection{Mean Regression Results\label{mean-regression-results}}

Results from mean regressions with \PARPMTEN as the only air pollutant are
presented in Table \ref{tab:mainthree}. Columns 1 and 3 show the Equation \eqref{eq:one} estimates.
Columns 2 and 4 show the Equation \eqref{eq:two} estimates, which include
interaction terms between maternal education and environmental exposure
variables. In columns 1 and 2, extreme temperature is defined by 1\% extreme tails of past temperatures, while columns 3 and 4 use
2.5\% as the tail cutoffs.

After controlling for conception year-by-month fixed effects, \PARPMTEN,
extreme cold, and extreme heat are all negatively and substantially
associated with birth weight. Column 1 in Table \ref{tab:mainthree} shows that a 1 $\mu g/m^3$
increase in average daily \PARPMTEN potential exposure during pregnancy is
associated with an 17.8 gram reduction of birth weight (s.e. 2.3). An
additional percentage point increase in potential exposure to extreme
heat or cold is associated with a 22.4 gram (s.e. 9.4) or 30.8 gram (s.e. 9.8)
birth weight reduction, respectively. Stated differently, a one
standard deviation change in potential exposure to \PARPMTEN (i.e., 6.6
$\mu g/m^3$), extreme heat (i.e., 1.61\% of pregnant days), or extreme cold (i.e., 1.00\% of pregnant days) corresponds to 0.25, 0.08, and 0.07 standard
deviation changes in birth weights, respectively.\footnote{While the
  magnitudes of the associations between \PARPMTEN or other pollutants and
  birth weights in relatively low pollution settings are generally smaller
  than what we find, our results are similar to the findings
  obtained by \textcite{bharadwaj_atmospheric_2008} from Chile---a setting with
  similar mean \PARPMTEN exposures as our setting here---where results show that
  a one standard deviation (17 $\mu g/m^3$) increase in \PARPMTEN exposures is
  associated with 0.23 standard deviations (125 grams) of birth-weight
  reduction.} The results from Columns 1 and 3 are similar. The
exception is that in Column 3, with the less-extreme temperature
thresholds, heat exposure is no longer significantly negatively associated with birth
weight.

Columns 2 and 4 show the maternal college education interaction coefficient estimates.\footnote{The coefficient estimates for college education appear negative in the interaction specifications (columns 2 and 4). In these specifications, college education coefficients represent intercepts (adjusting for other variables in the model) for the hypothetical situation of no \PARPMTEN, cold, or heat exposure. The predicted protective effects of maternal education given the observed range of ambient environmental measures are presented in our discussion of college premiums for birth weights.}
 In both columns, mothers' college education
status is associated with reduced vulnerability to ambient air pollution
and extreme heat exposures. However, this protective effect of 
mothers' college is not significant for extreme cold. Focusing on Column 2, there
is a positive college-education interaction coefficient of 2.0 (s.e.
0.7) for \PARPMTEN and 11.8 (s.e. 2.9) for extreme heat. Comparing
college-educated to non-college-educated mothers, the interactions
correspond to a 10\% dampening of the negative association of \PARPMTEN
with birth weight and a 44\% dampening of the negative association
of extreme heat with birth weight.

We evaluate the college premium in birth weight, which we define to be the predicted
gap in birth weight between college-educated and non-college-educated
mothers, by considering the negative intercept for college education,
-115.7 (s.e. 53.2), and the positive education slopes jointly. Holding
extreme heat and cold exposures at their respective means, the college
premium amounts to 0.05, 0.08, and 0.16 standard deviations of birth
weight at the 1\textsuperscript{st} (63 $\mu g/m^3$), 50\textsuperscript{th}
(70 $\mu g/m^3$), and 99\textsuperscript{th} (89 $\mu g/m^3$) percentiles of the
marginal distribution of \PARPMTEN. Similarly, holding \PARPMTEN and extreme cold
at their respective means, the college premium amounts to 0.05, 0.07 and
0.17 standard deviations of birth weight at the 1\textsuperscript{st} (0\%), 50\textsuperscript{th} (0.76\%) and
99\textsuperscript{th} (4.71\%) percentiles of the marginal
distribution of the percentage of pregnancy days exposed to extreme heat.

\hypertarget{quantile-regression-results}{%
\subsection{Quantile Regression
Results}\label{quantile-regression-results}}

Conditional quantile results, without and with maternal education
interactions with the three environmental measures,
are shown in Tables \ref{tab:mainfour} and \ref{tab:mainfive}, respectively. These estimates use the more
stringent cold and heat exposure measures defined by the 1\%
cutoffs. Because these are conditional quantile estimates, the quantiles are defined with respect to what we call the children's unobserved innate vulnerabilities (i.e., the residuals in the birth weight regressions, not the birth weights themselves). 

\hypertarget{baseline-quantile-regression-results}{%
\subsubsection{\texorpdfstring{Baseline Quantile Regression Results
}{Baseline Quantile Regression Results }}\label{baseline-quantile-regression-results}}

Table \ref{tab:mainfour} shows a negative gradient for college education over ascending
conditional quantile estimates, indicating that the protective effects of college
education on birth weight decrease as we move from the left to the right tail of the conditional distribution of birth weight. Specifically, the coefficient estimates on mothers'
college completion status are 64.6 (s.e. 8.1), 21.8 (s.e. 4.8), and 11.2
(s.e. 7.4) at the 10\textsuperscript{th}, 50\textsuperscript{th}, and
90\textsuperscript{th} conditional quantiles. These amount to 0.14, 0.05,
and 0.02 standard deviations of birth weight.

Additionally, the negative associations between the three environmental
measures and birth weight are all stronger at lower conditional
quantiles. The \PARPMTEN coefficients at the 10\textsuperscript{th},
50\textsuperscript{th}, and 90\textsuperscript{th} conditional quantiles
are -38.9 (s.e. 3.5), -22.2 (s.e. 2.1), and -17.5 (s.e. 2.8). A one
standard deviation increase in potential \PARPMTEN exposures is associated
with 0.55, 0.31, and 0.25 standard deviation reductions of birth weight,
respectively. For extreme heat and cold, respectively, the magnitudes of the negative associations are
2.4 and 3.7 times larger at the 10\textsuperscript{th} conditional
quantile versus the 90.\textsuperscript{th} At the
10\textsuperscript{th} conditional quantiles, a one standard deviation
increase in extreme heat or cold is associated with a 0.06 or 0.13
standard deviation reduction of birth weights, respectively.

Table \ref{tab:mainfour} also indicates increasing estimates for males, decreasing
estimates for mother's age, as well as increasing estimates for mother's
age squared along ascending conditional quantiles estimates. The
directions and magnitudes of these coefficient estimates are similar to those
reported in the existing research on conditional and 
unconditional quantile estimates for birth weights \autocite{abrevaya_effects_2001, koenker_quantile_2001, firpo_unconditional_2009}.

\hypertarget{quantile-regression-results-with-interactions}{%
\subsubsection{Quantile Regression Results with
Interactions}\label{quantile-regression-results-with-interactions}}

At lower conditional quantiles, maternal college education ameliorates the negative
associations between birth weight and pollution and extreme heat, but not
extreme cold. As shown in Table \ref{tab:mainfive} and Figure \ref{fig:mainthree}, the college-education
interaction coefficient estimates for \PARPMTEN are 5.7 (s.e. 1.3), 1.3
(s.e. 0.7), and 0.0 (s.e. 1.1) at the 10\textsuperscript{th},
50\textsuperscript{th}, and 90\textsuperscript{th} conditional
quantiles. These respectively correspond to 14, 6, and 0\%
reductions in the substantial negative associations between \PARPMTEN and birth
weight compared to non-college-educated mothers. Additionally, the
college-education interaction coefficient estimates for extreme heat are
26.4 (s.e. 5.4), 7.3 (s.e. 3.3), and 6.2 (s.e. 4.8) at the
10\textsuperscript{th}, 50\textsuperscript{th}, and
90\textsuperscript{th} conditional quantiles. While the overall associations of extreme heat with birth weights are less substantial than the associations of \PARPMTEN with birth weights, these coefficients, respectively,
correspond to  71, 31, and 47\% reductions in the associations between extreme heat and birth weight.

Considering jointly the positive slopes and negative intercepts from Table \ref{tab:mainfive}, the college
premium in birth weight is magnified at lower conditional quantiles.
Holding heat and cold exposure variables at their means, the college premiums at
the 1\textsuperscript{st}, 50\textsuperscript{th} and
99\textsuperscript{th} percentiles of the marginal distribution of \PARPMTEN
amount to 0.02, 0.02, and 0.02 standard deviations of birth weight with
90\textsuperscript{th} conditional quantile estimates, and are magnified
substantially to 0.03, 0.11, and 0.34 standard deviations of birth weight
with 10\textsuperscript{th} conditional quantile estimates. These
results are visualized in Figure \ref{fig:mainfour}. Additionally, holding \PARPMTEN and cold
exposures at their means, the college premiums at the
1\textsuperscript{st}, 50\textsuperscript{th} and 99\textsuperscript{th}
percentiles of the marginal distribution of extreme heat amount to 0.00,
0.01, and 0.06 standard deviations of birth weight with the
90\textsuperscript{th} conditional quantile estimates, and are magnified
to 0.06, 0.10, and 0.32 standard deviations with the
10\textsuperscript{th} conditional quantile estimates. In contrast, the associations between extreme cold exposures
and birth weights do not vary across maternal education groups, so there are no significant variations in the college premium in birth weight as extreme cold exposures increase.

\subsection{Composite Index Analysis\label{composite-index-analysis}}

We repeat the analyses parallel to that for \PARPMTEN but with the composite index pollution
measure. Detailed results are presented in Appendix Tables \ref{tab:appone}, \ref{tab:apptwo}, and \ref{tab:appthree}. Patterns
in the temperature results are robust to the choice of pollution measure,
and are not discussed here. The results from the composite index
analysis closely mirror the patterns revealed in the \PARPMTEN analysis.
There is a protective effect of maternal education with regard to
pollution, and the finding persists across conditional quantiles of
birth weight. Further, quantile
regressions show that the negative associations between pollution
and birth weight, and the protective effects of maternal education with
respect to pollution, are magnified at lower conditional quantiles.

A key difference in the composite index analysis results is that for estimates based on Equation \eqref{eq:one} without maternal education
interactions, the birth weight change associated with one standard
deviation change in the composite index is 1.8 times of the corresponding
change in the \PARPMTEN analysis. In the regressions with interactions, the
protective effect of college education is weaker in the
composite index analysis than in the \PARPMTEN analysis. Together, these
factors lead to a pattern in which the college premium in grams of birth
weight along the marginal distribution of the composite index is
approximately the same as the college premium in birth weight along the
marginal distribution of \PARPMTEN.

\subsection{Summary}
Conditional on ambient temperature exposures, we find that one standard deviation increase in ambient \PARPMTEN is associated with a 0.25 standard deviation reduction in birth weight. At average ambient \PARPMTEN and extreme temperature exposure levels, the protective effect of maternal college education is up to 0.07 standard deviations of birth weight. We also find that the negative association between ambient exposures and birth weight is two times as large at lower conditional quantiles of birth weight than at the median. At lower conditional quantiles of birth weight, a one standard deviation
increase in ambient \PARPMTEN exposure is associated with a reduction in birth weight of up to 0.48 standard
deviations. 

Additionally, the protection associated with university-educated mothers with respect to pollution and extreme heat is five times as large at the lowest decile of conditional quantiles of birth weight than at the highest decile. In particular, the protective effect of college
education can be up to 0.34 standard deviations of birth weight when
children with higher unobserved innate vulnerabilities---those at
lower conditional quantiles---face high levels of ambient \PARPMTEN and
extreme temperature exposures.  Though we focus on a continuous measure of birth weight for all terms in the text, we find similar results with binary measures used in some other studies for low birth weight, preterm births, and small for gestational age (Appendix Section \ref{sec:appbinary}). Our results show that there are substantively significant heterogeneous moderating effects of maternal education on the negative associations between ambient environmental exposures and birth weight. We are the first to document these relationships using mean and conditional quantile regressions with an approach in which we consider three key ambient environmental exposures jointly. 

\section{Discussion and Conclusion \label{summary-and-conclusions}}

In this paper, we link a database of all births recorded for three years
born in hospitals in a district in Guangzhou, China, to daily measures of particulate
matter (\PARPMTEN), nitrogen dioxide, and sulfur dioxide and to daily
meteorological measurements during pregnancies. Using conditional mean
and quantile regressions, we estimate the associations of birth weight
with air pollution, measured by ambient \PARPMTEN and a composite pollution
measure, and with extreme cold and hot temperatures. To our knowledge, this paper is the first study of birth outcomes in a fairly high-pollution context to consider effect modifications of both pollution and extreme weather by maternal education. It is also novel in addressing whether the protective effects of
maternal education are heterogeneous with respect to babies' unobserved innate vulnerabilities. 

Our approach has certain limitations.  First, we cannot observe mothers' addresses, which would have enabled us to match to the air pollution level from the closest monitoring station. On the other hand, our approach of using core city districts' pollution levels is reasonable if we assume that women who give birth in a district are likely to live and move around primarily in that district and neighboring districts. We are encouraged that our results are not sensitive to our choice of a proxy for ambient exposure: results are robust to whether we use city-wide pollution measures or center-city district pollution measures (Appendix Section \ref{sec:appallcity}). 

A second limitation of the paper is that it is geographically limited to one locale. Under the assumption of homogeneous environmental effects across locations, more cities would be beneficial in terms of additional variance in environmental and temperature exposure variables and would improve the precision of estimates. However, if effects are heterogeneous across locations, an average effect obtained from information over multiple locations might not properly capture varying local effects. Accounting for differences across locations appropriately is potentially especially important for temperature effects, which are a focus in our paper. Relative temperature matters---the implications for pregnancy of a given temperature may differ across temperature zones.  Focusing on a single region has the advantage of avoiding this problem.\footnote{Our paper capitalizes on full administrative data, rather than sample data from one location. For a study that is focused on distributional outcomes, it is important to capture the full range of birth weight distributions and we are able to do this.  There are benefits to having more people in the same location, which allows for careful control of location-specific time-trends that could vary with changes in local economic conditions that might correlate with birth weight.} The experience of Guangzhou may not generalize across China, but it offers a case study that is potentially relevant to other urban settings in south China, and possibly parts of Southeast Asia, characterized by broadly similar climates, high levels of pollution, and economic conditions. Finally, the administrative data to which we have access has excellent measurement of birth outcomes, relative to self-reported survey data.  However, it contains very limited information about mothers. This limitation precludes analysis of other dimensions of socioeconomic status beyond maternal education or behavioral mechanisms that might explain the patterns we find.

With these caveats in mind, we find strong negative associations between ambient \PARPMTEN and extreme
temperature exposures and birth weights. For example, a one standard
deviation increase in \PARPMTEN exposure---6.6 $\mu g/m^3$---is
associated with a 116 to 258 gram reduction in birth weights. One
standard deviation---on average, 2.7 days---increase in prenatal exposure
to extreme cold temperatures (lower than -0.87 degrees Celsius in universal
thermal climate indices) is associated with a 17 to 63 gram reduction in
birth weights. One standard deviation---on average, 4.3
days---increase in prenatal exposure to extreme hot temperatures (higher
than 34.23 degrees Celsius, in universal thermal climate indices) is associated
with a 12 to 31 gram reduction in birth weights. Stronger associations are
seen for children with greater levels of unobserved innate vulnerabilities at lower conditional quantiles.

There is also strong evidence for effect modification with maternal
education. Newborns of mothers with college attainment experience an amelioration of
effects on birth weight of \PARPMTEN by about 0 to 14\%, and of extreme
hot weather by 31 to 70\%, with larger buffering effects experienced by those at lower conditional quantiles. However, there is no effect modification with maternal education in response to extreme cold
weather. This result is understandable in a context in which
centralized heating is not widely available. Although there is usually only a short
spell of very cold weather in each year, our results indicate that even
a few days of very cold weather exposure is associated with lower birth
weights. Given the lack of indoor heating, even mothers with higher
socioeconomic status might have no effective way of protecting
themselves from exposure to extreme cold weather. In contrast, air
conditioning was widely available during the time period of this study
and was very accessible for higher socioeconomic status mothers.

In short, our findings demonstrate that babies of college-educated
mothers---compared to babies of other mothers in the same district of
Guangzhou---experienced lower risk in the context of air
pollution and extreme heat exposure. It is important to note that the
benefits accruing to babies of college-educated mothers vary. The protective effects are most pronounced among those
experiencing more environmental risk---more extreme ambient pollution
and heat exposure. Protective effects of maternal education also emerge
more strongly for infants with greater unobserved innate
vulnerabilities. In other words, socioeconomic disparities
and underlying child health vulnerabilities both stratify the realized
impacts associated with a common ambient environment---especially when
that common ambient environment is extreme pollution or extreme hot temperatures.
Our findings suggest that babies at greatest risk from high pollution
levels and increasingly frequent heat events are those at the nexus of
socioeconomic and health vulnerability.

\pagebreak

\begingroup
\setstretch{1.0}
\setlength\bibitemsep{3pt}
\printbibliography[title=References]
\endgroup
\pagebreak

\newcommand{\TITLETABMAINONE}{Summary statistics.\label{tab:mainone}}
\newcommand{\TITLETABMAINONELONG}{\TITLETABMAINONE}
\newcommand{\TITLETABMAINTWO}{Distribution of cumulative ambient \PARPMTEN, composite index and extreme temperature exposures during the course of pregnancy.\label{tab:maintwo}}
\newcommand{\TITLETABMAINTHREE}{OLS regression analysis of birth weight with interactions of ambient \PARPMTEN and extreme temperature with maternal education.\label{tab:mainthree}}
\newcommand{\TITLETABMAINTHREEALLCITY}{OLS regression analysis of birth weight with interactions of ambient \PARPMTEN and extreme temperature with maternal education. Using all city districts pollution measurements.\label{tab:mainthreeallcity}}
\newcommand{\TITLETABMAINTHREEFULLTERM}{OLS regression analysis of birth weight with interactions of ambient \PARPMTEN and extreme temperature with maternal education. Using only full-term births.\label{tab:mainthreefullterm}}
\newcommand{\TITLETABMAINTHREEBIMARGIN}{Marginal-effects from logistic regression analysis of binary birth outcomes with interactions of ambient \PARPMTEN and extreme temperature with maternal education.\label{tab:mainthreebimargin}}
\newcommand{\TITLETABMAINTHREEBIODDS}{Odds-ratio from Logistic regression analysis of binary birth outcomes with interactions of ambient \PARPMTEN and extreme temperature with maternal education.\label{tab:mainthreebiodds}}
\newcommand{\TITLETABMAINFOUR}{Conditional quantile regression analysis of birth weight with ambient \PARPMTEN and extreme temperature and maternal education.\label{tab:mainfour}}
\newcommand{\TITLETABMAINFIVE}{Conditional quantile regression analysis of birth weight with interactions of ambient \PARPMTEN and extreme temperature with maternal education.\label{tab:mainfive}}
\newcommand{\TITLETABAPPONE}{OLS regression analysis of birth weight with interactions of ambient composite index and extreme temperature with maternal education.\label{tab:appone}}
\newcommand{\TITLETABAPPTWO}{Conditional quantile regression analysis of birth weight with ambient composite index and extreme temperature and maternal education.\label{tab:apptwo}}
\newcommand{\TITLETABAPPTHREE}{Conditional quantile regression analysis of birth weight with interactions of ambient composite index and extreme temperature with maternal education.\label{tab:appthree}}
\newcommand{\TITLEFIGONE}{Daily mean ambient air pollution levels for three major pollutants, 2008 to 2011, Guangzhou, China.\label{fig:mainone}}
\newcommand{\TITLEFIGONETEMP}{
Daily mean temperature (Universal Thermal Climate Index), Guangzhou, China, 2008 to 2011.\label{fig:mainonetemp}
}
\newcommand{\TITLEFIGTWO}{Daily mean ambient \PARPMTEN exposure by mother's education.\label{fig:maintwo}}
\newcommand{\TITLEFIGTHREE}{Graphical illustration of mean and conditional quantile estimation coefficients from models of birth weight (grams) for  average daily mean ambient \PARPMTEN exposures ($\mu g/m^3$) (top row) and composite index (bottom row) for mothers with college education and mothers with high school or less education (HS).\label{fig:mainthree}}
\newcommand{\TITLEFIGFOUR}{Graphical illustration of predicted college premium in birth weight for different conditional quantiles and at different ambient \PARPMTEN exposure levels.\label{fig:mainfour}}

\newcommand{\EMPHNOTES}{\emph{Note:}\thinspace}
\newcommand{\CONTROLS}{\EMPHNOTES Regressions control for conception year and month interaction fixed effects, day of the week at birth, parity, and daily mean rainfall.}
\newcommand{\SIGESTIEACH}{$^{\dagger}$ $p<0.10$; $\sym{*}$ $p<0.05$; $\sym{**}$ $p<0.01$.}
\newcommand{\SIGESTIEACHQUANT}{\SIGESTIEACH Bootstrap standard errors are shown in parentheses.}
\newcommand{\SIGQUANTCOMP}{\\Given bootstrapped simultaneous conditional quantile estimates, superscripts a, b, c, and d indicate whether estimates across conditional quantiles are statistically different for the three ambient environment or education variables.
\\\textsuperscript{a} P10, P25, and P50 are significantly different at 0.05 sig. level.
\\\textsuperscript{b} P10 and P90 are significantly different at 0.05 sig. level.
\\\textsuperscript{c} P25 and P75 are significantly different at 0.05 sig. level.
\\\textsuperscript{d} P50, P75, and P90 are significantly different at 0.05 sig. level.}
\newcommand{\TABNOTESMAINTABONE}{\small \textit{Note: } The analytic sample has 53,879 observations. The college subsample has 17,967 observations and the high school or lower subsample has 35,912 observations. Sources and computations for ambient environmental variables are discussed in the data section of the paper.\\
$^{\ddagger}$ P-value from testing whether the mean gap for each variable between college-educated and non-college-educated mothers is statistically different.}
\newcommand{\TABNOTESMAINTABTWO}{\small \textit{Note: }Figures reported in the table represent distributional statistics along the marginal distribution of ambient pollution and extreme temperature variables.}
\newcommand{\TABNOTESMAINTABTHREE}{\small\CONTROLS\\\SIGESTIEACH}
\newcommand{\TABNOTESMAINTABTHREEALLCITY}{\small\CONTROLS\\\SIGESTIEACH}
\newcommand{\TABNOTESMAINTABTHREEFULLTERM}{\small\CONTROLS\\\SIGESTIEACH}
\newcommand{\TABNOTESMAINTABTHREEBIMARGIN}{\small\CONTROLS\\\SIGESTIEACH}
\newcommand{\TABNOTESMAINTABTHREEBIODDS}{\small\CONTROLS\\\SIGESTIEACH}
\newcommand{\TABNOTESMAINTABFOUR}{\small\CONTROLS\SIGQUANTCOMP\\\SIGESTIEACHQUANT}
\newcommand{\TABNOTESMAINTABFIVE}{\small\CONTROLS\SIGQUANTCOMP\\\SIGESTIEACHQUANT}
\newcommand{\TABNOTESAPPTABONE}{\small\CONTROLS\\\SIGESTIEACH}
\newcommand{\TABNOTESAPPTABTWO}{\small\CONTROLS\SIGQUANTCOMP\\\SIGESTIEACHQUANT}
\newcommand{\TABNOTESAPPTABTHREE}{\small\CONTROLS\SIGQUANTCOMP\\\SIGESTIEACHQUANT}
\newcommand{\FIGNOTESMAINFIGONE}{\small\EMPHNOTES This figure shows daily means for three monitored ambient air pollutants in Guangzhou between 2008 and 2011.}
\newcommand{\FIGNOTESMAINFIGONETEMP}{\small
}
\newcommand{\FIGNOTESMAINFIGTWO}{\small\EMPHNOTES The left panel contains observed data, while the right panel depicts the conditional distribution of daily mean ambient \PARPMTEN exposure after controlling for conception year and month interaction fixed effects.}
\newcommand{\FIGNOTESMAINFIGTHREE}{\small\EMPHNOTES The x-axis of the sub-plots corresponds to results from different conditional quantile estimations. 
We interpret lower conditional quantiles as corresponding to higher levels of unobserved innate vulnerabilities for babies. 
The two subfigures in the top panel are based on estimates from Tables \ref{tab:mainthree} and \ref{tab:mainfive}.
The two subfigures in the bottom panel are based on estimates from Appendix Tables \ref{tab:appone} and \ref{tab:appthree}. 
The y-axis reports regression coefficients. The left and right panels depict the same information in two ways.
The solid line in the right panel shows the gap between the HS and college quantile coefficient lines in the corresponding left panel. The gaps depicted on the right provide a visualization of the maternal college education and pollution interaction coefficients, from Table \ref{tab:mainfive} in the top panel for \PARPMTEN or from Table \ref{tab:appthree} for the composite index.
In the top panel, coefficients indicate grams of birth weight change for each additional $\mu g/m^3$ of average daily mean ambient \PARPMTEN exposures. In the bottom panel, coefficients indicate grams of birth weight change for each additional index unit of the composite index.}
\newcommand{\FIGNOTESMAINFIGFOUR}{\small\EMPHNOTES The x-axis of the figure represents different ambient \PARPMTEN exposure levels during the course of pregnancy, which correspond to the range of ambient \PARPMTEN exposure levels reported in Table \ref{tab:maintwo}. The y-axis reports predicted college premium in birth weight--the predicted gap in birth weight between college-educated and non-college-educated mothers. Given estimates from Tables \ref{tab:mainthree} and \ref{tab:mainfive}, each line corresponds to predictions for children at differing conditional quantiles. The premium is greater at higher ambient exposure levels and for children with higher levels of unobserved innate vulnerabilities--those at lower conditional quantiles.}


\pagebreak 
\begin{figure}[H]
	\centering
	\caption{\TITLEFIGONE}
	\includegraphics[width=1.0\textwidth, center]{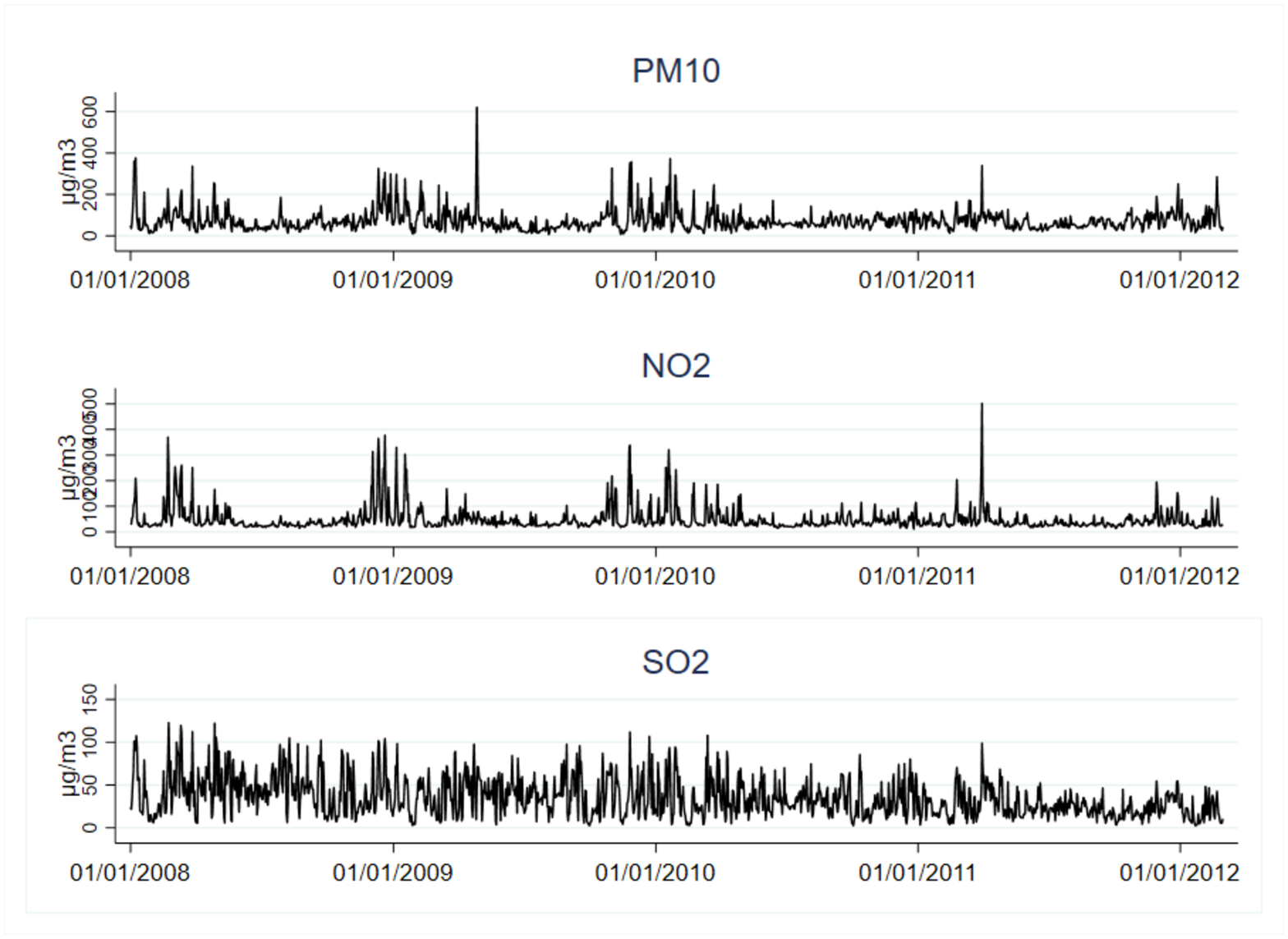}
	\captionsetup{width=1.0\textwidth}\caption*{\FIGNOTESMAINFIGONE}
\end{figure}
\pagebreak 
\begin{figure}[H]
	\centering
	\caption{\TITLEFIGONETEMP}
	\includegraphics[width=0.75\textwidth, center]{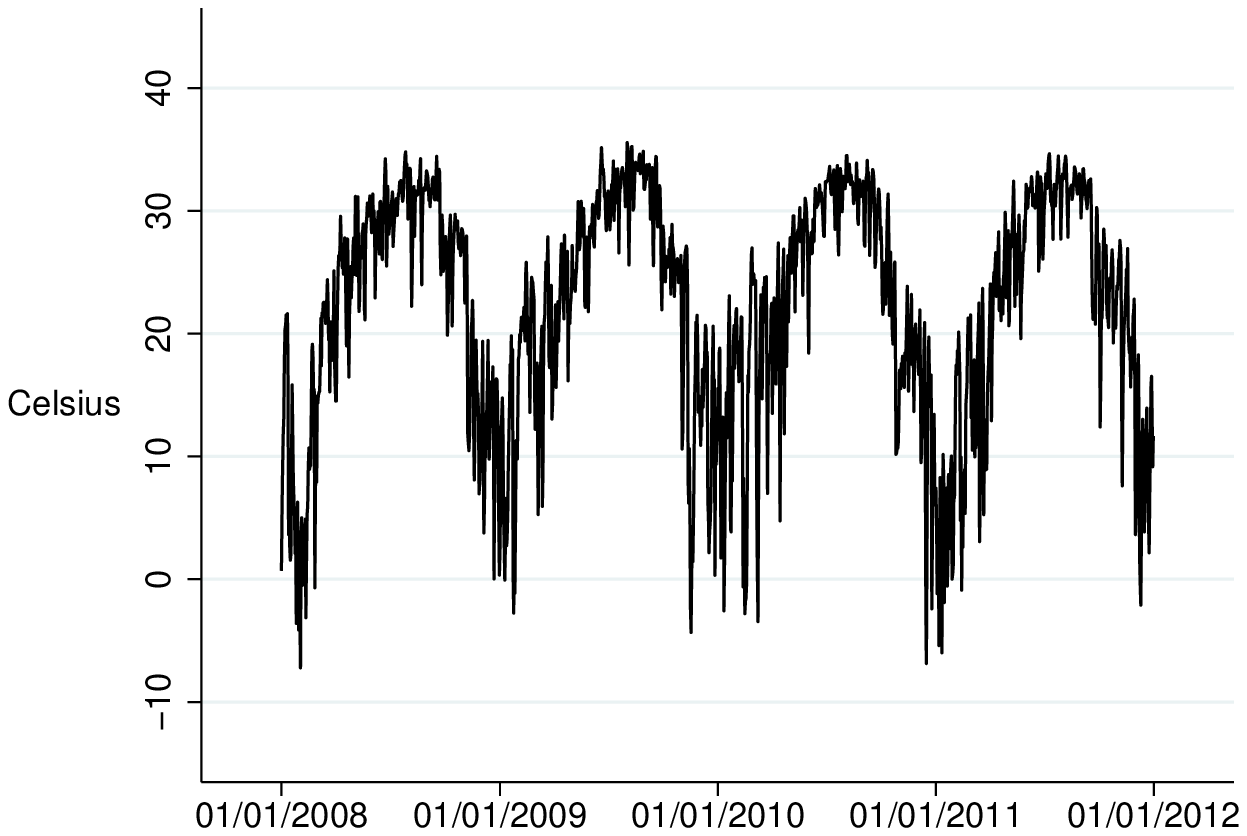}
	\captionsetup{width=0.75\textwidth}\caption*{\FIGNOTESMAINFIGONETEMP}
\end{figure}
\pagebreak 
\begin{figure}[H]
	\centering
	\caption{\TITLEFIGTWO}
	\includegraphics[width=1.0\textwidth, center]{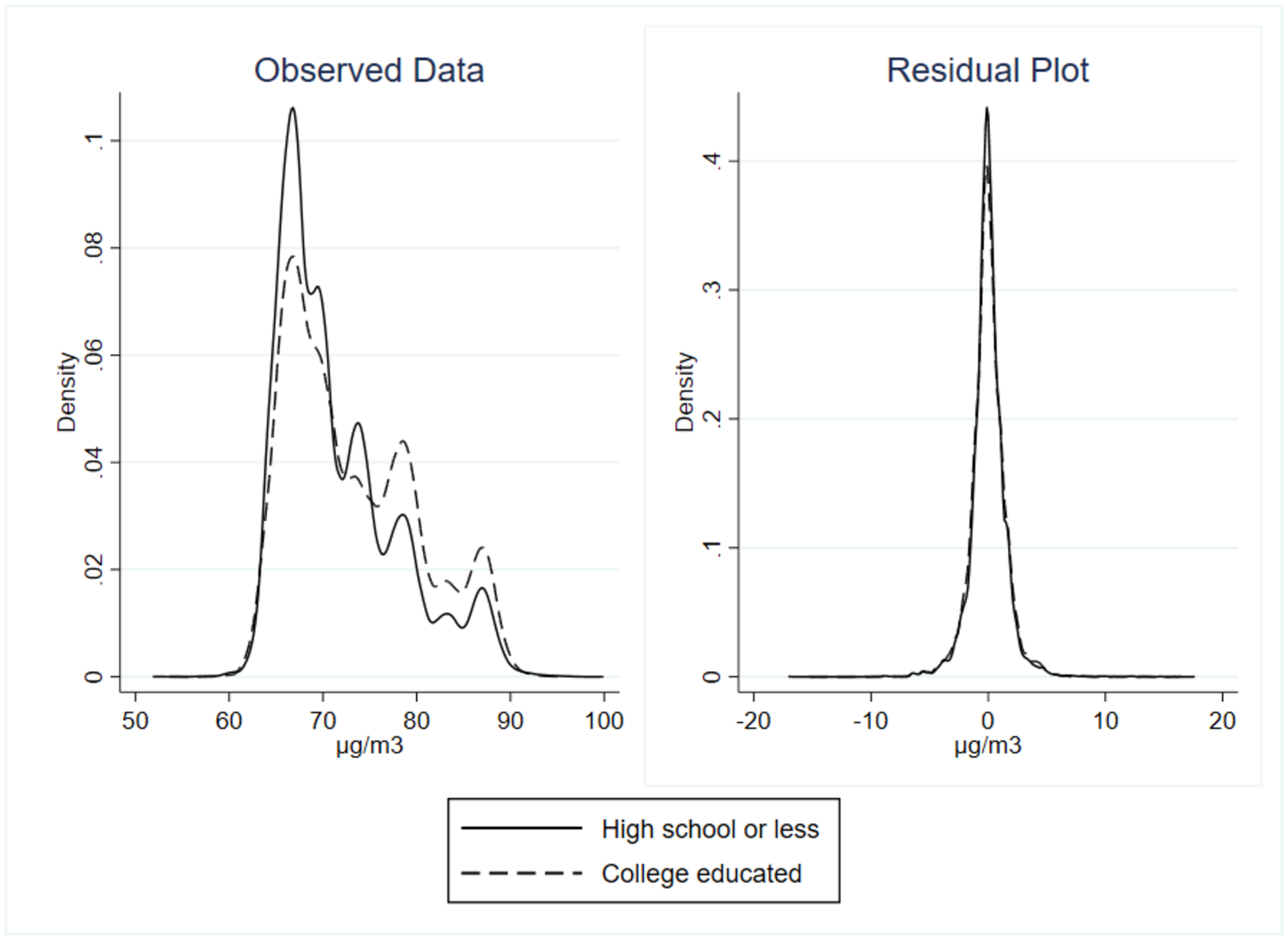}
	\captionsetup{width=1.0\textwidth}\caption*{\FIGNOTESMAINFIGTWO}
\end{figure}
\pagebreak 
\begin{figure}[H]
	\centering
	\captionsetup{width=1.00\textwidth}\caption{\TITLEFIGTHREE}
	\includegraphics[width=1.00\textwidth, center]{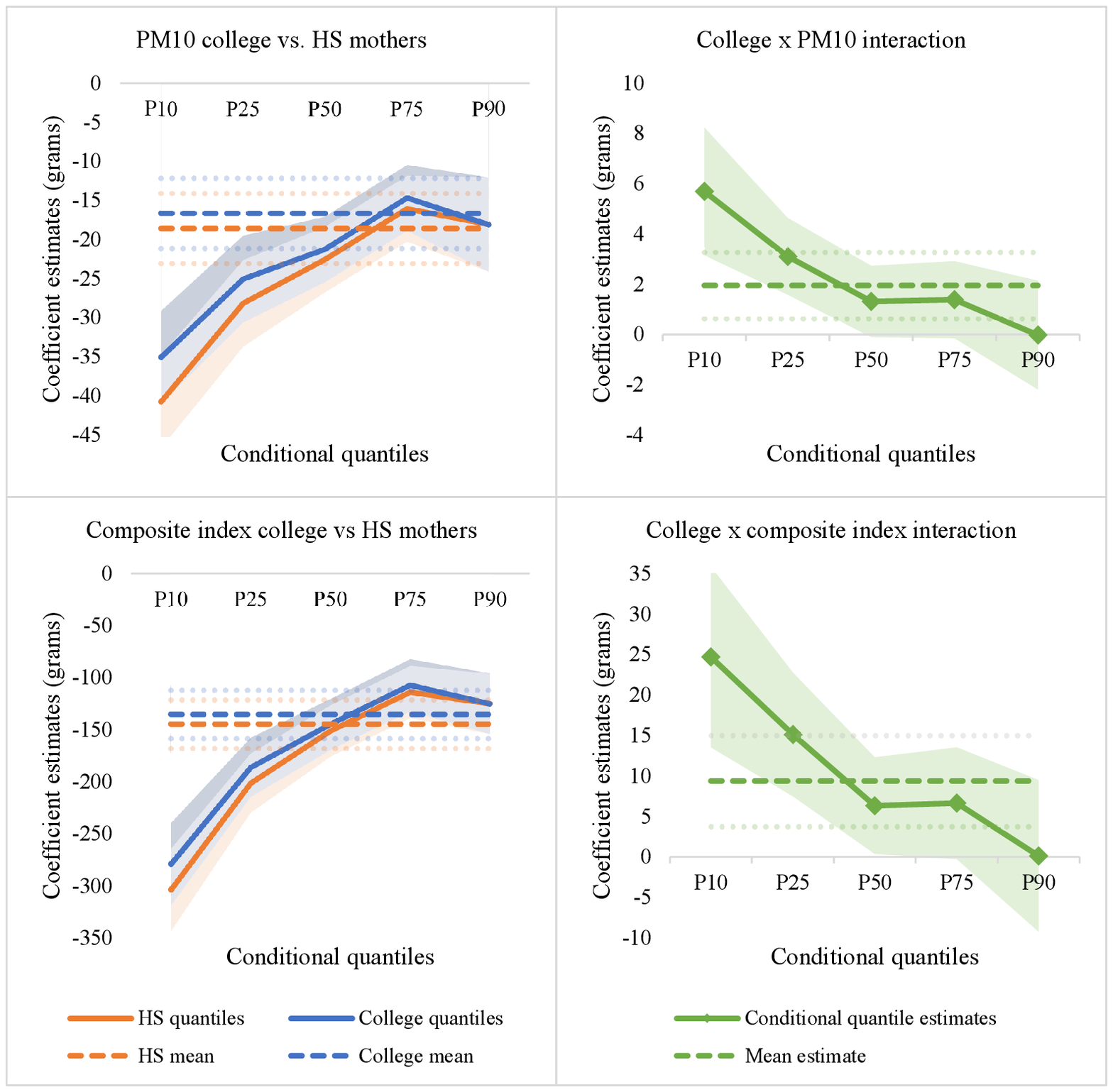}
	\captionsetup{width=1.00\textwidth}\caption*{\FIGNOTESMAINFIGTHREE}
\end{figure}
\pagebreak 
\begin{figure}[H]
	\centering
	\caption{\TITLEFIGFOUR}
	\includegraphics[width=1.0\textwidth, center]{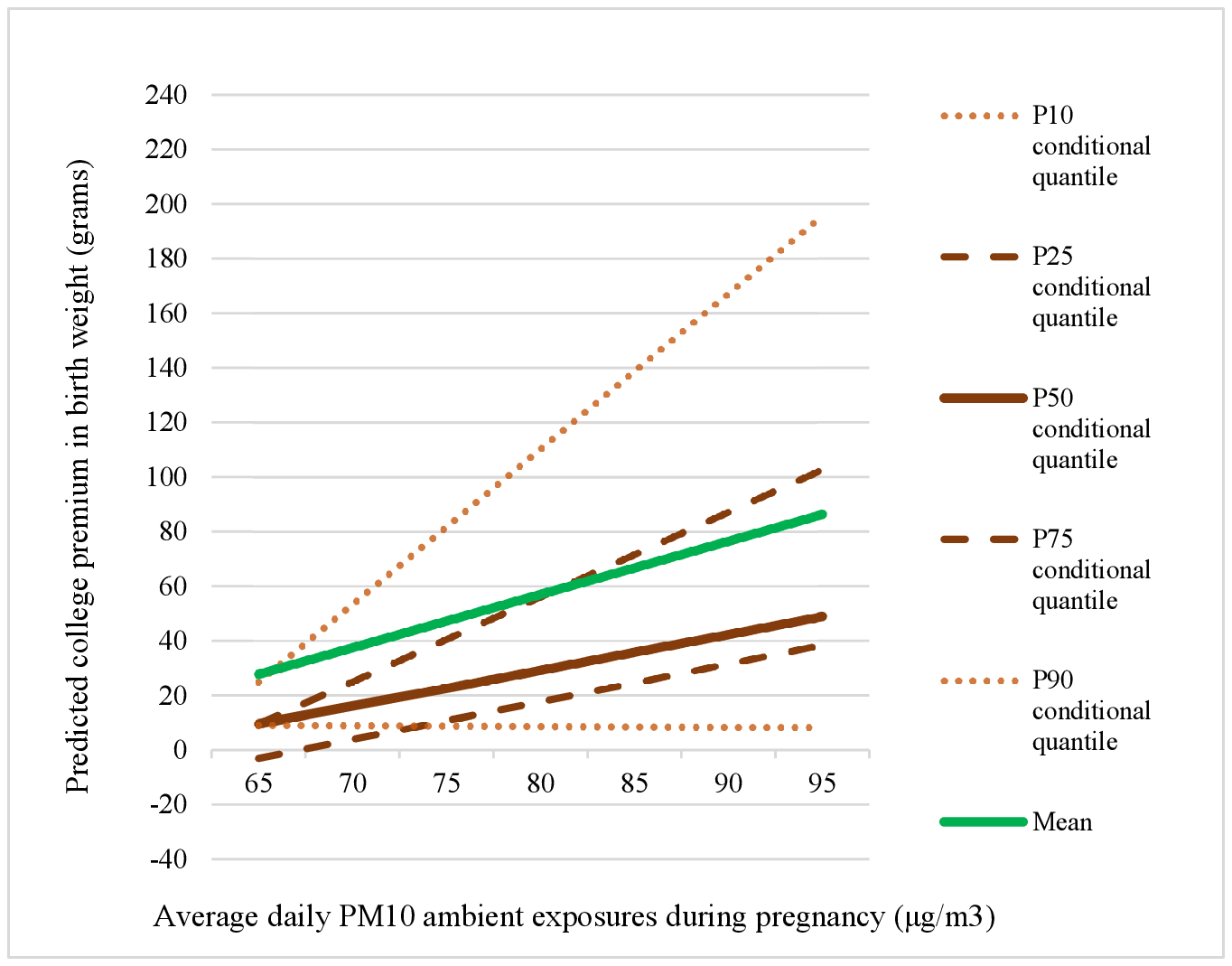}
	\captionsetup{width=1.0\textwidth}\caption*{\FIGNOTESMAINFIGFOUR}
\end{figure}
\clearpage


\pagebreak 
\begin{table}[htbp]
\centering
\caption{\hspace*{0mm}\TITLETABMAINONELONG}
\begin{adjustbox}{max width=1.0\textwidth}
\begin{tabular}{m{7.75cm}*{7}{>{\centering\arraybackslash}m{1.5cm}}}
\toprule
& & & \multicolumn{5}{c}{Comparison by mother's education}\\
\cmidrule(l{5pt}r{5pt}){4-8}
& \multicolumn{2}{c}{All} & \multicolumn{2}{c}{<= High school} & \multicolumn{2}{c}{>=College} & \multicolumn{1}{c}{Gap}\\
\cmidrule(l{5pt}r{5pt}){2-3} \cmidrule(l{5pt}r{5pt}){4-5} \cmidrule(l{5pt}r{5pt}){6-7} \cmidrule(l{5pt}r{5pt}){8-8} 
 & mean & s.d. & mean & s.d. & mean & s.d. & p-value \\
\midrule
\addlinespace
\multicolumn{8}{l}{\hspace*{0mm}Child variables}\\
\addlinespace
\hspace*{6mm}Sex (male=1) & 0.53 & 0.50 & 0.54 & 0.50 & 0.53 & 0.50 & (0.01)\\
\addlinespace
\hspace*{6mm}Birth weight (grams) & 3182 & 473 & 3167 & 492 & 3213 & 429 & (0.00)\\
\addlinespace
\hspace*{6mm}Gestational age (days) & 273 & 11.51 & 273 & 12.31 & 273 & 9.69 & (0.00)\\
\addlinespace
\hspace*{6mm}Low birth weight    &  0.06 &        0.23&        0.07&        0.25&        0.04&        0.19&      (0.00)\\
\addlinespace
\hspace*{6mm}Preterm (\%, gestational age $<$ 37weeks) & 0.07 & 0.26 & 0.08 & 0.27 & 0.05 & 0.23 & (0.00)\\
\addlinespace
\hspace*{6mm}Small for Gestational Age   &        0.09&        0.29&        0.10&        0.30&        0.08&        0.27&      (0.00)\\
\addlinespace
\addlinespace
\multicolumn{8}{l}{\hspace*{0mm}Maternal variables}\\
\addlinespace
\hspace*{6mm}Mother's age (years) & 29.08 & 4.17 & 28.84 & 4.51 & 29.57 & 3.35 & (0.00)\\
\addlinespace
\hspace*{6mm}Mother's schooling attainment (years) & 13.13 & 2.19 & 11.75 & 0.90 & 15.89 & 1.15 & (0.00)\\
\addlinespace
\hspace*{6mm}Parity  & 1.25 & 0.51 & 1.34 & 0.57 & 1.06 & 0.26 & (0.00)\\
\addlinespace
\addlinespace
\multicolumn{8}{l}{\hspace*{0mm}Average of daily mean potential pollution exposures during pregnancy }\\
\addlinespace
\addlinespace
\multicolumn{8}{l}{\hspace*{6mm}\textit{All city districts measurements}}\\
\addlinespace
\hspace*{6mm}\PARPMTEN ($\mu g/m^3$) & 73.22 & 7.16 & 72.52 & 6.89 & 74.64 & 7.49 & (0.00)\\
\addlinespace
\hspace*{6mm}NO\textsubscript{2} ($\mu g/m^3$) & 41.41 & 6.13 & 40.61 & 6.00 & 43.01 & 6.07 & (0.00)\\
\addlinespace
\hspace*{6mm}SO\textsubscript{2} ($\mu g/m^3$) & 35.18 & 4.63 & 34.46 & 4.64 & 36.62 & 4.26 & (0.00)\\
\addlinespace
\hspace*{6mm}Composite index of \PARPMTEN, NO\textsubscript{2}, SO\textsubscript{2} & -0.05 & 1.56 & -0.26 & 1.53 & 0.38 & 1.53 & (0.00)\\
\addlinespace
\addlinespace
\multicolumn{8}{l}{\hspace*{6mm}\textit{Center-city districts measurements}}\\
\addlinespace
\hspace*{6mm}\PARPMTEN ($\mu g/m^3$) & 72.10 & 6.63 & 71.53 & 6.36 & 73.23 & 6.99 & (0.00)\\
\addlinespace
\hspace*{6mm}NO\textsubscript{2} ($\mu g/m^3$) & 48.60 & 6.94 & 47.73 & 6.79 & 50.32 & 6.90 & (0.00)\\
\addlinespace
\hspace*{6mm}SO\textsubscript{2} ($\mu g/m^3$) & 34.44 & 4.62 & 33.74 & 4.68 & 35.83 & 4.18 & (0.00)\\
\addlinespace
\hspace*{6mm}Composite index of \PARPMTEN, NO\textsubscript{2}, SO\textsubscript{2} & -0.02 & 1.48 & -0.22 & 1.45 & 0.37 & 1.46 & (0.00)\\
\addlinespace
\addlinespace
\multicolumn{8}{l}{\hspace*{0mm}Temperature and rainfall during pregancy}\\
\addlinespace
\hspace*{6mm}Average daily rainfall (mm) & 5.15 & 1.61 & 5.20 & 1.64 & 5.03 & 1.55 & (0.00)\\
\addlinespace
\hspace*{6mm}Average daily mean temperature (C$^\circ$) & 22.38 & 1.83 & 22.32 & 1.86 & 22.50 & 1.77 & (0.00)\\
\addlinespace
\addlinespace
\multicolumn{8}{l}{\hspace*{6mm}\textit{Percent of pregnancy days with potential exposure to extreme temperatures}}\\
\addlinespace
\hspace*{6mm}Extreme heat, above past 99\% & 1.73 & 1.61 & 1.60 & 1.54 & 1.98 & 1.71 & (0.00)\\
\addlinespace
\hspace*{6mm}Extreme cold, below past 1\% & 1.70 & 1.00 & 1.77 & 1.01 & 1.57 & 0.95 & (0.00)\\
\addlinespace
\hspace*{6mm}Extreme heat, above past 97.5\%  & 3.73 & 2.87 & 3.54 & 2.75 & 4.11 & 3.08 & (0.00)\\
\addlinespace
\hspace*{6mm}Extreme cold, below past 2.5\% & 4.40 & 2.47 & 4.53 & 2.52 & 4.13 & 2.33 & (0.00)\\
\addlinespace
\bottomrule
\addlinespace[0.5em]
\multicolumn{8}{p{1.35\textwidth}}{\parbox[t]{1.35\textwidth}{\TABNOTESMAINTABONE}}\\
\end{tabular}
\end{adjustbox}
\end{table}

\pagebreak 
\begin{table}[htbp]
\centering
\captionsetup{width=.90\textwidth}
\caption{\hspace*{0mm}\TITLETABMAINTWO}
\begin{adjustbox}{max width=0.9\textwidth}
\begin{tabular}{m{2.5cm}*{6}{>{\centering\arraybackslash}m{2cm}}}
\toprule
& & & \multicolumn{4}{c}{Varying cutoffs for extreme temperature exposures}\\
\cmidrule(l{5pt}r{5pt}){4-7}
& \multicolumn{2}{c}{Pollution measures} & \multicolumn{2}{c}{1 percent cutoff} & \multicolumn{2}{c}{2.5 percent cutoff}\\
\cmidrule(l{5pt}r{5pt}){2-3} \cmidrule(l{5pt}r{5pt}){4-5} \cmidrule(l{5pt}r{5pt}){6-7}
Statistics & \PARPMTEN & composite & heat & cold & heat & cold \\
       & {\footnotesize$\mu g/m^3$} & {\footnotesize index} & {\footnotesize percent days} & {\footnotesize percent days} & {\footnotesize percent days} & {\footnotesize percent days} \\
\midrule
\addlinespace
\multicolumn{7}{l}{\hspace*{0mm}Percentiles}\\
\addlinespace
\hspace*{6mm}P1 & 63.05  & -2.31 & 0.00 & 0.00 & 0.00 & 0.00\\
\addlinespace
\hspace*{6mm}P5 & 64.50  & -2.18 & 0.00 & 0.36 & 0.00 & 0.37\\
\addlinespace
\hspace*{6mm}P10 & 65.10 & -1.71 & 0.00 & 0.43 & 0.38 & 1.08\\
\addlinespace
\hspace*{6mm}P25 & 66.80 & -1.33 & 0.71 & 0.73 & 1.75 & 2.50\\
\addlinespace
\hspace*{6mm}P50 & 70.01 & -0.25 & 0.76 & 1.79 & 2.59 & 4.23\\
\addlinespace
\hspace*{6mm}P75 & 76.41 & 1.20  & 3.65 & 2.63 & 6.88 & 6.83\\
\addlinespace
\hspace*{6mm}P90 & 82.56 & 2.29  & 4.38 & 2.93 & 8.39 & 7.33\\
\addlinespace
\hspace*{6mm}P95 & 86.45 & 2.66  & 4.48 & 3.00 & 8.58 & 7.49\\
\addlinespace
\hspace*{6mm}P99 & 88.52 & 2.92  & 4.71 & 3.17 & 9.02 & 7.94\\
\addlinespace
\addlinespace
\multicolumn{7}{l}{\hspace*{0mm}Min and Max}\\
\addlinespace
\hspace*{6mm}Min & 51.86 & -2.42 & 0.00 & 0.00 & 0.00 & 0.00\\
\addlinespace
\hspace*{6mm}Max & 99.91 & 4.86 & 6.35 & 4.35 & 12.17 & 10.87\\
\addlinespace
\bottomrule
\addlinespace[0.5em]
\multicolumn{7}{p{1.03\textwidth}}{\parbox[t]{1.03\textwidth}{\TABNOTESMAINTABTWO}}\\
\end{tabular}
\end{adjustbox}
\end{table}

\pagebreak 
\begin{table}[htbp]
\centering
\captionsetup{width=0.9\textwidth}
\caption{\hspace*{0mm}\TITLETABMAINTHREE}
\begin{adjustbox}{max width=0.9\textwidth}
\begin{tabular}{m{5.7cm}*{4}{>{\centering\arraybackslash}m{2cm}}}
\toprule
& \multicolumn{4}{c}{Varying cutoffs of extreme temperature exposures}\\
\cmidrule(l{5pt}r{5pt}){2-5}
& \multicolumn{2}{c}{1 percent cutoff} & \multicolumn{2}{c}{2.5 percent cutoff}\\
\cmidrule(l{5pt}r{5pt}){2-3} \cmidrule(l{5pt}r{5pt}){4-5} 
Variable & (1) & (2) & (3) & (4) \\
\midrule
\addlinespace
\multicolumn{5}{l}{\hspace*{0mm}Environmental exposure variables}\\
\addlinespace
\hspace*{6mm}\PARPMTEN & -17.83\sym{**} & -18.58\sym{**} & -14.59\sym{**} & -15.41\sym{**}\\
\addlinespace
 & (2.27) & (2.29) & (2.31) & (2.33)\\
\addlinespace
\hspace*{6mm}Extreme heat & -22.38\sym{*} & -27.06\sym{**} & 0.09 & -2.33\\
\addlinespace
 & (9.41) & (9.57) & (5.11) & (5.21)\\
\addlinespace
\hspace*{6mm}Extreme cold & -30.84\sym{**} & -30.33\sym{**} & -24.68\sym{**} & -24.33\sym{**}\\
\addlinespace
 & (9.80) & (9.96) & (4.37) & (4.43)\\
\addlinespace
\addlinespace
\multicolumn{5}{l}{\hspace*{0mm}Education and environmental exposure interactions}\\
\addlinespace
\hspace*{6mm}College educated & 44.55\sym{**} & -115.70\sym{*} & 44.85\sym{**} & -133.00\sym{*}\\
\addlinespace
 & (4.43) & (53.17) & (4.43) & (55.52)\\
\addlinespace
\hspace*{6mm}College x \PARPMTEN &  & 1.95\sym{**} &  & 2.19\sym{**}\\
\addlinespace
 &  & (0.68) &  & (0.70)\\
\addlinespace
\hspace*{6mm}College x extreme heat &  & 11.82\sym{**} &  & 6.26\sym{**}\\
\addlinespace
 &  & (2.93) &  & (1.71)\\
\addlinespace
\hspace*{6mm}College x extreme cold &  & -2.15 &  & -1.35\\
\addlinespace
 &  & (4.68) &  & (1.87)\\
\addlinespace
\addlinespace
\multicolumn{5}{l}{\hspace*{0mm}Control variables}\\
\addlinespace
\hspace*{6mm}Male & 104.00\sym{**} & 104.00\sym{**} & 103.90\sym{**} & 103.90\sym{**}\\
\addlinespace
 & (3.93) & (3.93) & (3.93) & (3.93)\\
\addlinespace
\hspace*{6mm}Mother’s age & 55.41\sym{**} & 54.90\sym{**} & 55.33\sym{**} & 54.85\sym{**}\\
\addlinespace
 & (4.98) & (4.98) & (4.98) & (4.98)\\
\addlinespace
\hspace*{6mm}Mother’s age$^2$ & -0.90\sym{**} & -0.89\sym{**} & -0.89\sym{**} & -0.89\sym{**}\\
\addlinespace
 & (0.08) & (0.08) & (0.08) & (0.08)\\
\addlinespace
\hspace*{0mm}Intercept & 1,205.00\sym{**} & 1,276.00\sym{**} & 1,019.00\sym{**} & 1,094.00\sym{**}\\
\addlinespace
 & (219.90) & (221.50) & (221.40) & (223.30)\\
\addlinespace
\midrule
Observations & 53,879 & 53,879 & 53,879 & 53,879\\
R$^2$	& 0.069 & 0.069 & 0.069 & 0.070\\
\bottomrule
\addlinespace[0.5em]
\multicolumn{5}{p{0.95\textwidth}}{\parbox[t]{0.95\textwidth}{\TABNOTESMAINTABTHREE}}\\
\end{tabular}
\end{adjustbox}
\end{table}

\pagebreak 
\begin{table}[htbp]
\centering
\captionsetup{width=1.0\textwidth}
\caption{\hspace*{0mm}\TITLETABMAINFOUR}
\begin{adjustbox}{max width=1.0\textwidth}
\begin{tabular}{m{5.7cm}*{5}{>{\centering\arraybackslash}m{2cm}}}
\toprule
& \multicolumn{5}{c}{Estimates at conditional quantiles}\\
\cmidrule(l{5pt}r{5pt}){2-6} 
Variable & P10 & P25 & P50 & P75 & P90\\
\midrule
\addlinespace
\multicolumn{6}{l}{\hspace*{0mm}Environmental exposure variables}\\
\addlinespace
\hspace*{6mm}\PARPMTEN $^{a,b,c,d}$ & -38.90\sym{**} & -26.34\sym{**} & -22.16\sym{**} & -15.30\sym{**} & -17.50\sym{**}\\
\addlinespace
 & (3.49) & (2.44) & (2.09) & (2.38) & (2.83)\\
\addlinespace
\hspace*{6mm}Extreme heat & -17.88 & -13.15$^\dagger$ & -19.06\sym{*} & -8.10 & -7.43\\
\addlinespace
 & (13.26) & (7.84) & (8.90) & (11.00) & (12.17)\\
\addlinespace
\hspace*{6mm}Extreme cold $^{a,b}$ & -62.61\sym{**} & -29.79\sym{**} & -19.68\sym{*} & -23.89\sym{*} & -16.82\\
\addlinespace
 & (16.30) & (9.85) & (9.37) & (10.15) & (13.97)\\
\addlinespace
\addlinespace
\multicolumn{6}{l}{\hspace*{0mm}Education}\\
\addlinespace
\hspace*{6mm}College educated $^{a,b,c,d}$ & 64.59\sym{**} & 33.20\sym{**} & 21.79\sym{**} & 9.10 & 11.19\\
\addlinespace
 & (8.08) & (5.72) & (4.81) & (5.87) & (7.39)\\
\addlinespace
\addlinespace
\multicolumn{6}{l}{\hspace*{0mm}Control variables}\\
\addlinespace
\hspace*{6mm}Male & 92.62\sym{**} & 98.92\sym{**} & 111.94\sym{**} & 117.30\sym{**} & 132.60\sym{**}\\
\addlinespace
 & (7.67) & (5.07) & (4.42) & (4.80) & (6.75)\\
\addlinespace
\hspace*{6mm}Mother’s age & 64.30\sym{**} & 45.91\sym{**} & 46.47\sym{**} & 51.18\sym{**} & 46.37\sym{**}\\
\addlinespace
 & (10.01) & (5.79) & (5.63) & (5.53) & (8.81)\\
\addlinespace
\hspace*{6mm}Mother’s age$^2$ & -1.09\sym{**} & -0.73\sym{**} & -0.73\sym{**} & -0.78\sym{**} & -0.69\sym{**}\\
\addlinespace
 & (0.17) & (0.10) & (0.09) & (0.09) & (0.15)\\
\addlinespace
\hspace*{0mm}Intercept & 917.89\sym{*} & 1,168.25\sym{**} & 1,835.52\sym{**} & 1,897.93\sym{**} & 2,226.62\sym{**}\\
\addlinespace
 & (378.70) & (252.98) & (200.27) & (232.69) & (334.94)\\
\addlinespace
\midrule
Observations & 53,879 & 53,879 & 53,879 & 53,879 & 53,879\\
\bottomrule
\addlinespace[0.5em]
\multicolumn{6}{p{1.09\textwidth}}{\parbox[t]{1.09\textwidth}{\TABNOTESMAINTABFOUR}}\\
\end{tabular}
\end{adjustbox}
\end{table}

\pagebreak 
\begin{table}[htbp]
\centering
\captionsetup{width=1.0\textwidth}
\caption{\hspace*{0mm}\TITLETABMAINFIVE}
\begin{adjustbox}{max width=1.0\textwidth}
\begin{tabular}{m{5.7cm}*{5}{>{\centering\arraybackslash}m{2cm}}}
\toprule
& \multicolumn{5}{c}{Estimates at conditional quantiles}\\
\cmidrule(l{5pt}r{5pt}){2-6} 
Variable & P10 & P25 & P50 & P75 & P90\\
\midrule
\addlinespace
\multicolumn{6}{l}{\hspace*{0mm}Environmental exposure variables}\\
\addlinespace
\hspace*{6mm}\PARPMTEN $^{a,b,c,d}$ & -40.75\sym{**} & -28.15\sym{**} & -22.55\sym{**} & -16.04\sym{**} & -18.06\sym{**}\\
\addlinespace
 & (3.03) & (2.83) & (2.14) & (2.15) & (3.09)\\
\addlinespace
\hspace*{6mm}Extreme heat & -37.30\sym{**} & -23.23\sym{*} & -23.50\sym{**} & -12.04 & -13.12\\
\addlinespace
 & (13.73) & (10.64) & (8.23) & (10.92) & (13.12)\\
\addlinespace
\hspace*{6mm}Extreme cold $^{a,b}$ & -65.43\sym{**} & -28.98\sym{**} & -17.15\sym{*} & -22.28\sym{*} & -13.74\\
\addlinespace
 & (15.79) & (10.57) & (8.75) & (10.45) & (12.38)\\
\addlinespace
\addlinespace
\multicolumn{6}{l}{\hspace*{0mm}Education and environmental exposure interactions}\\
\addlinespace
\hspace*{6mm}College educated $^{a,b,c}$ & -396.40\sym{**} & -216.76\sym{**} & -83.69 & -95.51 & 2.20\\
\addlinespace
 & (99.62) & (59.02) & (58.77) & (60.23) & (87.96)\\
\addlinespace
\hspace*{6mm}College x \PARPMTEN $^{a,b}$ & 5.69\sym{**} & 3.11\sym{**} & 1.31$^\dagger$ & 1.39$^\dagger$ & -0.02\\
\addlinespace
 & (1.30) & (0.78) & (0.72) & (0.78) & (1.10)\\
\addlinespace
\hspace*{6mm}College x extreme heat $^{a,b}$ & 26.35\sym{**} & 11.83\sym{**} & 7.30\sym{*} & 5.24 & 6.19\\
\addlinespace
 & (5.42) & (3.91) & (3.27) & (3.50) & (4.75)\\
\addlinespace
\hspace*{6mm}College x extreme cold & 3.30 & 2.21 & -2.68 & -3.97 & -1.34\\
\addlinespace
 & (9.12) & (6.28) & (4.94) & (5.87) & (7.69)\\
\addlinespace
\addlinespace
\multicolumn{6}{l}{\hspace*{0mm}Control variables}\\
\addlinespace
\hspace*{6mm}Male & 90.14\sym{**} & 98.32\sym{**} & 112.02\sym{**} & 116.40\sym{**} & 132.55\sym{**}\\
\addlinespace
 & (8.38) & (5.15) & (4.42) & (5.15) & (6.12)\\
\addlinespace
\hspace*{6mm}Mother’s age & 62.49\sym{**} & 44.68\sym{**} & 46.26\sym{**} & 48.33\sym{**} & 45.30\sym{**}\\
\addlinespace
 & (10.02) & (6.04) & (5.07) & (5.70) & (8.91)\\
\addlinespace
\hspace*{6mm}Mother’s age$^2$ & -1.06\sym{**} & -0.71\sym{**} & -0.73\sym{**} & -0.74\sym{**} & -0.67\sym{**}\\
\addlinespace
 & (0.17) & (0.10) & (0.09) & (0.10) & (0.15)\\
\addlinespace
\hspace*{0mm}Intercept & 1,240.85\sym{**} & 1,415.85\sym{**} & 1,888.54\sym{**} & 2,032.87\sym{**} & 2,357.66\sym{**}\\
\addlinespace
 & (371.67) & (263.70) & (206.46) & (202.99) & (369.89)\\
\addlinespace
\midrule
Observations & 53,879 & 53,879 & 53,879 & 53,879 & 53,879\\
\bottomrule
\addlinespace[0.5em]
\multicolumn{6}{p{1.09\textwidth}}{\parbox[t]{1.09\textwidth}{\TABNOTESMAINTABFIVE}}\\
\end{tabular}
\end{adjustbox}
\end{table}

\pagebreak
\clearpage

\appendix

\setlength{\footnotemargin}{5.75mm}
\begingroup
\doublespacing
\centering
\Large ONLINE APPENDIX \\
\Large\begin{singlespace}\href{\PAPERDOIURL}{\PAPERTITLE}\end{singlespace}
\large \AUTHORLIU, 
\AUTHORBEHRMAN, 
\AUTHORHANNUM, 
\AUTHORWANG, 
and \AUTHORZHAO\\[1.0em]
\endgroup

\renewcommand{\thefigure}{A.\arabic{figure}}
\setcounter{figure}{0}
\renewcommand{\thetable}{A.\arabic{table}}
\setcounter{table}{0}
\renewcommand{\theequation}{A.\arabic{equation}}
\setcounter{equation}{0}
\renewcommand{\thefootnote}{A.\arabic{footnote}}
\setcounter{footnote}{0}

\section{Additional Estimation Results}

\subsection{Composite Pollution Index}
In this Appendix section, we repeat the analyses for \PARPMTEN with the composite index pollution
measure. Results are presented in Appendix Tables \ref{tab:appone}, \ref{tab:apptwo} and \ref{tab:appthree}. Patterns
in the temperature results are robust to choice of pollution measure. The results from the composite index
analysis closely mirror the patterns revealed in the \PARPMTEN analysis.
There is a protective effect of maternal education with regard to
pollution, and the finding persists across conditional quantiles of
birth weight (see the dotted lines in Figure \ref{fig:mainthree} Bottom Row). Further, quantile
regressions show that the negative associations between the pollution
and birth weight, and the protective effects of maternal education with
respect to pollution, are magnified at lower conditional quantiles (see
the solid lines and shaded areas in Figure \ref{fig:mainthree} Bottom Row). See Section \ref{composite-index-analysis} for more discussion.

\begin{table}[htbp]
\centering
\captionsetup{width=0.9\textwidth}
\caption{\hspace*{0mm}\TITLETABAPPONE}
\begin{adjustbox}{max width=0.9\textwidth}
\begin{tabular}{m{5.7cm}*{4}{>{\centering\arraybackslash}m{2cm}}}
\toprule
& \multicolumn{4}{c}{Varying cutoffs of extreme temperature exposures}\\
\cmidrule(l{5pt}r{5pt}){2-5}
& \multicolumn{2}{c}{1 percent cutoff} & \multicolumn{2}{c}{2.5 percent cutoff}\\
\cmidrule(l{5pt}r{5pt}){2-3} \cmidrule(l{5pt}r{5pt}){4-5} 
Variable & (1) & (2) & (3) & (4) \\
\midrule
\addlinespace
\multicolumn{5}{l}{\hspace*{0mm}Environmental exposure variables}\\
\addlinespace
\hspace*{6mm}Composite index & -140.90\sym{**} & -144.50\sym{**} & -127.10\sym{**} & -131.00\sym{**}\\
\addlinespace
 & (11.82) & (11.90) & (11.88) & (11.95)\\
\addlinespace
\hspace*{6mm}Extreme heat & -28.61\sym{**} & -32.21\sym{**} & -4.47 & -6.28\\
\addlinespace
 & (9.49) & (9.64) & (5.06) & (5.16)\\
\addlinespace
\hspace*{6mm}Extreme cold & -27.97\sym{**} & -27.43\sym{**} & -23.38\sym{**} & -23.04\sym{**}\\
\addlinespace
 & (9.76) & (9.92) & (4.36) & (4.42)\\
\addlinespace
\addlinespace
\multicolumn{5}{l}{\hspace*{0mm}Education and environmental exposure interactions}\\
\addlinespace
\hspace*{6mm}College educated & 44.61\sym{**} & 29.45\sym{**} & 44.89\sym{**} & 29.82\sym{*}\\
\addlinespace
 & (4.43) & (11.15) & (4.43) & (12.06)\\
\addlinespace
\hspace*{6mm}College x composite index &  & 9.38\sym{**} &  & 10.43\sym{**}\\
\addlinespace
 &  & (2.87) &  & (2.84)\\
\addlinespace
\hspace*{6mm}College x extreme heat &  & 9.19\sym{**} &  & 4.73\sym{**}\\
\addlinespace
 &  & (2.77) &  & (1.57)\\
\addlinespace
\hspace*{6mm}College x extreme cold &  & -2.25 &  & -1.28\\
\addlinespace
 &  & (4.68) &  & (1.87)\\
\addlinespace
\addlinespace
\multicolumn{5}{l}{\hspace*{0mm}Control variables}\\
\addlinespace
\hspace*{6mm}Male & 104.30\sym{**} & 104.30\sym{**} & 104.20\sym{**} & 104.20\sym{**}\\
\addlinespace
 & (3.92) & (3.92) & (3.92) & (3.92)\\
\addlinespace
\hspace*{6mm}Mother’s age & 55.11\sym{**} & 54.54\sym{**} & 55.07\sym{**} & 54.51\sym{**}\\
\addlinespace
 & (4.97) & (4.97) & (4.97) & (4.97)\\
\addlinespace
\hspace*{6mm}Mother’s age$^2$ & -0.89\sym{**} & -0.88\sym{**} & -0.89\sym{**} & -0.88\sym{**}\\
\addlinespace
 & (0.08) & (0.08) & (0.08) & (0.08)\\
\addlinespace
\hspace*{0mm}Intercept & -218.50 & -203.60 & -147.60 & -133.10\\
\addlinespace
 & (136.30) & (136.40) & (136.80) & (136.90)\\
\addlinespace
\midrule
Observations & 53,879 & 53,879 & 53,879 & 53,879\\
R$^2$	& 0.071	& 0.072	& 0.072	& 0.072\\
\bottomrule
\addlinespace[0.5em]
\multicolumn{5}{p{0.95\textwidth}}{\parbox[t]{0.95\textwidth}{\TABNOTESAPPTABONE}}\\
\end{tabular}
\end{adjustbox}
\end{table}

\pagebreak 
\begin{table}[htbp]
\centering
\captionsetup{width=1.0\textwidth}
\caption{\hspace*{0mm}\TITLETABAPPTWO}
\begin{adjustbox}{max width=1.0\textwidth}
\begin{tabular}{m{5.7cm}*{5}{>{\centering\arraybackslash}m{2cm}}}
\toprule
& \multicolumn{5}{c}{Estimates at conditional quantiles}\\
\cmidrule(l{5pt}r{5pt}){2-6} 
Variable & P10 & P25 & P50 & P75 & P90\\
\midrule
\addlinespace
\multicolumn{6}{l}{\hspace*{0mm}Environmental exposure variables}\\
\addlinespace
\hspace*{6mm}Composite index $^{a,b,c,d}$ & -292.98\sym{**} & -192.96\sym{**} & -150.52\sym{**} & -111.98\sym{**} & -123.83\sym{**}\\
\addlinespace
 & (19.50) & (14.10) & (11.00) & (12.61) & (17.25)\\
\addlinespace
\hspace*{6mm}Extreme heat & -38.50\sym{**} & -20.23\sym{*} & -21.52\sym{**} & -9.73 & -7.79\\
\addlinespace
 & (14.26) & (9.76) & (8.01) & (9.38) & (14.53)\\
\addlinespace
\hspace*{6mm}Extreme cold $^{a,b}$ & -58.07\sym{**} & -25.52\sym{**} & -14.43$^\dagger$ & -21.52\sym{*} & -12.97\\
\addlinespace
 & (16.27) & (9.62) & (8.16) & (10.45) & (12.83)\\
\addlinespace
\addlinespace
\multicolumn{6}{l}{\hspace*{0mm}Education}\\
\addlinespace
\hspace*{6mm}College educated $^{a,b,c,d}$ & 66.01\sym{**} & 33.64\sym{**} & 23.00\sym{**} & 9.63 & 9.46\\
\addlinespace
 & (8.56) & (5.72) & (4.85) & (6.11) & (7.50)\\
\addlinespace
\addlinespace
\multicolumn{6}{l}{\hspace*{0mm}Control variables}\\
\addlinespace
\hspace*{6mm}Male & 92.49\sym{**} & 99.08\sym{**} & 112.03\sym{**} & 117.86\sym{**} & 131.44\sym{**}\\
\addlinespace
 & (8.66) & (4.88) & (3.93) & (4.90) & (6.67)\\
\addlinespace
\hspace*{6mm}Mother’s age & 63.22\sym{**} & 45.16\sym{**} & 46.75\sym{**} & 49.19\sym{**} & 43.76\sym{**}\\
\addlinespace
 & (10.97) & (6.31) & (5.19) & (6.16) & (8.42)\\
\addlinespace
\hspace*{6mm}Mother’s age$^2$ & -1.07\sym{**} & -0.73\sym{**} & -0.73\sym{**} & -0.75\sym{**} & -0.64\sym{**}\\
\addlinespace
 & (0.19) & (0.10) & (0.09) & (0.10) & (0.14)\\
\addlinespace
\hspace*{0mm}Intercept & -1,626.65\sym{**} & -564.20\sym{**} & 260.17\sym{*} & 844.02\sym{**} & 1,028.68\sym{**}\\
\addlinespace
 & (228.99) & (164.71) & (128.89) & (163.68) & (256.45)\\
\addlinespace
\midrule
Observations & 53,879 & 53,879 & 53,879 & 53,879 & 53,879\\
\bottomrule
\addlinespace[0.5em]
\multicolumn{6}{p{1.09\textwidth}}{\parbox[t]{1.09\textwidth}{\TABNOTESAPPTABTWO}}\\
\end{tabular}
\end{adjustbox}
\end{table}

\pagebreak 
\begin{table}[htbp]
\centering
\captionsetup{width=1.0\textwidth}
\caption{\hspace*{0mm}\TITLETABAPPTHREE}
\begin{adjustbox}{max width=1.0\textwidth}
\begin{tabular}{m{5.7cm}*{5}{>{\centering\arraybackslash}m{2cm}}}
\toprule
& \multicolumn{5}{c}{Estimates at conditional quantiles}\\
\cmidrule(l{5pt}r{5pt}){2-6} 
Variable & P10 & P25 & P50 & P75 & P90\\
\midrule
\addlinespace
\multicolumn{6}{l}{\hspace*{0mm}Environmental exposure variables}\\
\addlinespace
\hspace*{6mm}Composite index $^{a,b,c,d}$ & -303.74\sym{**} & -201.06\sym{**} & -151.02\sym{**} & -113.65\sym{**} & -125.04\sym{**}\\
\addlinespace
 & (20.20) & (14.57) & (12.55) & (12.82) & (14.82)\\
\addlinespace
\hspace*{6mm}Extreme heat $^{a,b}$ & -53.19\sym{**} & -26.34\sym{*} & -23.50\sym{*} & -12.95 & -15.17\\
\addlinespace
 & (15.83) & (11.66) & (9.34) & (10.00) & (12.36)\\
\addlinespace
\hspace*{6mm}Extreme cold $^{a,b}$ & -63.15\sym{**} & -26.51\sym{*} & -13.36 & -19.48$^\dagger$ & -11.46\\
\addlinespace
 & (15.67) & (11.10) & (8.64) & (10.39) & (13.96)\\
\addlinespace
\addlinespace
\multicolumn{6}{l}{\hspace*{0mm}Education and environmental exposure interactions}\\
\addlinespace
\hspace*{6mm}College educated & 24.22 & 9.96 & 16.40 & 7.32 & -0.30\\
\addlinespace
 & (17.25) & (13.52) & (13.34) & (15.17) & (19.60)\\
\addlinespace
\hspace*{6mm}College x composite index $^{a,b,c}$ & 24.70\sym{**} & 15.12\sym{**} & 6.34\sym{*} & 6.63$^\dagger$ & 0.12\\
\addlinespace
 & (5.69) & (3.90) & (3.04) & (3.52) & (4.77)\\
\addlinespace
\hspace*{6mm}College x extreme heat $^{a}$ & 19.62\sym{**} & 9.08\sym{*} & 4.80 & 3.01 & 7.33$^\dagger$\\
\addlinespace
 & (5.59) & (3.60) & (3.01) & (3.44) & (4.04)\\
\addlinespace
\hspace*{6mm}College x extreme cold & 3.24 & 2.74 & -3.23 & -3.38 & -1.79\\
\addlinespace
 & (7.92) & (5.31) & (5.38) & (6.21) & (8.21)\\
\addlinespace
\addlinespace
\multicolumn{6}{l}{\hspace*{0mm}Control variables}\\
\addlinespace
\hspace*{6mm}Male & 88.50\sym{**} & 98.64\sym{**} & 111.87\sym{**} & 117.91\sym{**} & 131.37\sym{**}\\
\addlinespace
 & (8.14) & (4.92) & (4.50) & (5.04) & (6.36)\\
\addlinespace
\hspace*{6mm}Mother’s age & 64.58\sym{**} & 45.67\sym{**} & 46.16\sym{**} & 49.16\sym{**} & 44.13\sym{**}\\
\addlinespace
 & (10.77) & (5.67) & (4.72) & (6.09) & (8.70)\\
\addlinespace
\hspace*{6mm}Mother’s age$^2$ & -1.09\sym{**} & -0.73\sym{**} & -0.72\sym{**} & -0.75\sym{**} & -0.65\sym{**}\\
\addlinespace
 & (0.18) & (0.10) & (0.08) & (0.10) & (0.15)\\
\addlinespace
\hspace*{0mm}Intercept & -1,534.26\sym{**} & -562.73\sym{**} & 273.06\sym{*} & 845.59\sym{**} & 1,061.62\sym{**}\\
\addlinespace
 & (246.88) & (159.04) & (124.02) & (150.09) & (249.14)\\
\addlinespace
\midrule
Observations & 53,879 & 53,879 & 53,879 & 53,879 & 53,879\\
\bottomrule
\addlinespace[0.5em]
\multicolumn{6}{p{1.09\textwidth}}{\parbox[t]{1.09\textwidth}{\TABNOTESAPPTABTHREE}}\\
\end{tabular}
\end{adjustbox}
\end{table}

\clearpage

\subsection{All City Districts \PARPMTEN Measurements\label{sec:appallcity}}

Rather than using pollution data from monitoring stations in the four contiguous districts located in the center of the city, in this robustness check, we use pollution data from the whole city of Guangzhou, including information from districts that are further away. There are 11 districts in Guangzhou, five districts are located to the north of the center-city districts, and two districts are located to the south. Guangzhou city's total area is 7434 square kilometers. The four center-city districts' total area is 280 square kilometers. 

Given the proximity, it is more likely that mothers who deliver in the center-city district where our data comes from reside in center-city districts. However, it is also possible that some of them come from further-away districts. Additionally, in a metropolitan city with convenient local transport networks, it is likely that Guangzhou residents travel beyond their district borders throughout the city for work, leisure, shopping, and health and other social services. So it is plausible that the appropriate measure for ambient pollution should be based on city-wide data rather than district-specific data. 

Summary statistics in Table \ref{tab:mainone} shows that pollution measurements from all city districts and center-city districts are quiet similar. The mean and standard deviation of the average daily mean potential \PARPMTEN exposure are 73.2 and 7.2 for all city districts and 72.1 and 6.6 for center-city districts. The all city and center-city composite index also have similar means and standard deviations. The similarity in pollution exposure within the confines of Guangzhou is not surprising given that that these pollutants are formed and spread through lower-atmosphere and are usually not highly localized \autocite{he_severe_2019}. 

We present results using all-city district data in Table \ref{tab:mainthreeallcity}, which presents parallel results as in Table \ref{tab:mainthree}, except that center-city districts \PARPMTEN measurements are replaced by all city districts measurements. Table \ref{tab:mainthreeallcity} estimates are almost identical to Table \ref{tab:mainthree} results. Similarly, results from using the two types of measures are also similar for conditional quantile estimations and estimation using the composite index. Given their similarity, those results are not presented. 

\begin{table}[htbp]
\centering
\captionsetup{width=0.9\textwidth}
\caption{\hspace*{0mm}\TITLETABMAINTHREEALLCITY}
\begin{adjustbox}{max width=0.9\textwidth}
\begin{tabular}{m{5.7cm}*{4}{>{\centering\arraybackslash}m{2cm}}}
\toprule
& \multicolumn{4}{c}{Varying cutoffs of extreme temperature exposures}\\
\cmidrule(l{5pt}r{5pt}){2-5}
& \multicolumn{2}{c}{1 percent cutoff} & \multicolumn{2}{c}{2.5 percent cutoff}\\
\cmidrule(l{5pt}r{5pt}){2-3} \cmidrule(l{5pt}r{5pt}){4-5} 
Variable & (1) & (2) & (3) & (4) \\
\midrule
\addlinespace
\multicolumn{5}{l}{\hspace*{0mm}Environmental exposure variables}\\
\addlinespace
\hspace*{6mm}\PARPMTEN all city districts & -17.81\sym{**} & -18.55\sym{**} & -14.84\sym{**} & -15.64\sym{**}\\
\addlinespace
 & (2.18) & (2.21) & (2.21) & (2.24)\\
\addlinespace
\hspace*{6mm}Extreme heat & -21.91\sym{*} & -26.27\sym{**} & 0.15 & -2.10\\
\addlinespace
 & (9.51) & (9.67) & (5.15) & (5.25)\\
\addlinespace
\hspace*{6mm}Extreme cold & -33.55\sym{**} & -33.36\sym{**} & -25.75\sym{**} & -25.51\sym{**}\\
\addlinespace
 & (9.81) & (9.97) & (4.37) & (4.42)\\
\addlinespace
\addlinespace
\multicolumn{5}{l}{\hspace*{0mm}Education and environmental exposure interactions}\\
\addlinespace
\hspace*{6mm}College educated & 44.48\sym{**} & -117.60\sym{*} & 44.79\sym{**} & -132.60\sym{**}\\
\addlinespace
 & (4.43) & (48.99) & (4.43) & (49.90)\\
\addlinespace
\multicolumn{2}{l}{\hspace*{6mm}College x \PARPMTEN all city districts}  & 1.96\sym{**} &   & 2.07\sym{**}\\
\addlinespace
 &  & (0.61) &  & (0.61)\\
\addlinespace
\hspace*{6mm}College x extreme heat &  & 11.07\sym{**} &  & 5.83\sym{**}\\
\addlinespace
 &  & (2.85) &  & (1.64)\\
\addlinespace
\hspace*{6mm}College x extreme cold &  & -1.09 &  & -0.97\\
\addlinespace
 &  & (4.74) &  & (1.89)\\
\addlinespace
\addlinespace
\multicolumn{5}{l}{\hspace*{0mm}Control variables}\\
\addlinespace
\hspace*{6mm}Male & 104.00\sym{**} & 104.00\sym{**} & 103.90\sym{**} & 103.90\sym{**}\\
\addlinespace
 & (3.93) & (3.93) & (3.93) & (3.93)\\
\addlinespace
\hspace*{6mm}Mother’s age & 55.40\sym{**} & 54.87\sym{**} & 55.31\sym{**} & 54.81\sym{**}\\
\addlinespace
 & (4.98) & (4.98) & (4.98) & (4.98)\\
\addlinespace
\hspace*{6mm}Mother’s age$^2$ & -0.89\sym{**} & -0.89\sym{**} & -0.89\sym{**} & -0.89\sym{**}\\
\addlinespace
 & (0.08) & (0.08) & (0.08) & (0.08)\\
\addlinespace
\hspace*{0mm}Intercept & 1,216.00\sym{**} & 1,288.00\sym{**} & 1,019.00\sym{**} & 1,125.00\sym{**}\\
\addlinespace
 & (215.60) & (217.10) & (221.40) & (218.60)\\
\addlinespace
\midrule
Observations & 53,879 & 53,879 & 53,879 & 53,879\\
R$^2$	& 0.069 & 0.069 & 0.069 & 0.070\\
\bottomrule
\addlinespace[0.5em]
\multicolumn{5}{p{0.95\textwidth}}{\parbox[t]{0.95\textwidth}{\TABNOTESMAINTABTHREEALLCITY}}\\
\end{tabular}
\end{adjustbox}
\end{table}

\subsection{Sample Truncation Based on Gestational Age at Birth\label{sec:appfullterm}}

Research on the effects of pollution on birth weight often include both pre-term (with gestational age less than 37 weeks) and full-term births \autocite{stieb_ambient_2012, klepac_ambient_2018}. Some studies restrict the analytic sample to full-term births. Full-term sample selection may be important for creating a balanced sample to study the differential effects of pollution during different trimesters \autocite{morello-frosch_ambient_2010} or each month of gestation \autocite{huang_periconceptional_2020}. In this paper, we are interested in the average exposure effects from ambient pollution and extreme temperature throughout the course of pregnancy and do not estimate gestation-specific effects.\footnote{In Equations \eqref{eq:one} and \eqref{eq:two}, we implicitly assume that the effects of pollution and extreme temperature exposures are the same across gestational periods. Given our sample and location constraints, we do not have sufficient power to identify quantile- and gestation-period-specific estimates.} 

In the text, we focus on the distribution of birth weights. Following \textcite{abrevaya_effects_2001, koenker_quantile_2001, firpo_unconditional_2009}, we focus on all live singleton births and do not truncate based on gestational age. Gestational age in our sample ranges from 171 to 319 days. While we believe it is important to analyze the full distributional outcomes without gestational truncation, to facilitate comparisons to papers that do impose gestational truncation, in Table \ref{tab:mainthreefullterm}, we conduct the same estimation with the same variables as in Table \ref{tab:mainthree}, but now restrict our sample to only individuals with gestational age longer than 36 weeks. 

The main results in Table \ref{tab:mainthreefullterm}  are similar in directions but smaller in magnitudes than the results in Table \ref{tab:mainthree}: we continue to find negative associations between pollution and extreme temperatures and birth weight, and find that mothers' college education status is associated with reduced vulnerability to ambient air pollution and extreme heat exposures.\footnote{In both Tables \ref{tab:mainthree} and \ref{tab:mainthreefullterm}, we find a weak negative coefficient for the maternal-college-education and extreme-cold interaction variable. The coefficient is not significant in Table \ref{tab:mainthree} and weakly significant in Table \ref{tab:mainthreefullterm}. As shown in Table \ref{tab:mainfive}, these relationships fluctuate between weakly positive and negative across the quantiles for results without gestational restrictions. Quantile results with gestational restrictions are similar.} In particular, we find a 1 µg/m3 increase in average daily \PARPMTEN potential exposure during pregnancy is associated with an 8.85 gram reduction of birth weight (s.e. 1.56) in Table \ref{tab:mainthreefullterm} column 1, but the same increase is associated with a 17.83 gram reduction of birth weight (s.e. 2.27) in Table \ref{tab:mainthree} column 1. Additionally, we find a maternal college education and \PARPMTEN interaction coefficient of 0.72 (s.e. 0.59) in Table \ref{tab:mainthreefullterm} column 2 and of 1.95 (s.e. 0.68) in Table \ref{tab:mainthree} column 2. Corresponding quantile regression results with full-term birth also show similar patterns as in Tables 4 and 5. 

\begin{table}[htbp]
\centering
\captionsetup{width=0.9\textwidth}
\caption{\hspace*{0mm}\TITLETABMAINTHREEFULLTERM}
\begin{adjustbox}{max width=0.9\textwidth}
\begin{tabular}{m{5.7cm}*{4}{>{\centering\arraybackslash}m{2cm}}}
\toprule
& \multicolumn{4}{c}{Varying cutoffs of extreme temperature exposures}\\
\cmidrule(l{5pt}r{5pt}){2-5}
& \multicolumn{2}{c}{1 percent cutoff} & \multicolumn{2}{c}{2.5 percent cutoff}\\
\cmidrule(l{5pt}r{5pt}){2-3} \cmidrule(l{5pt}r{5pt}){4-5} 
Variable & (1) & (2) & (3) & (4) \\
\midrule
\addlinespace
\multicolumn{5}{l}{\hspace*{0mm}Environmental exposure variables}\\
\addlinespace
\hspace*{6mm}\PARPMTEN & -8.85\sym{**} & -9.14\sym{**} & -7.70\sym{**} & -8.05\sym{**}\\
\addlinespace
 & (1.56) & (1.58) & (1.57) & (1.60)\\
\addlinespace
\hspace*{6mm}Extreme heat & -8.56 & -10.37 & 0.15 & -0.71\\
\addlinespace
 & (6.68) & (6.78) & (3.52) & (3.58)\\
\addlinespace
\hspace*{6mm}Extreme cold & -16.54\sym{*} & -14.19\sym{*} & -12.69\sym{**} & -11.61\sym{**}\\
\addlinespace
 & (6.99) & (7.12) & (3.04) & (3.09)\\
\addlinespace
\addlinespace
\multicolumn{5}{l}{\hspace*{0mm}Education and environmental exposure interactions}\\
\addlinespace
\hspace*{6mm}College educated & 12.33\sym{**} & -35.51 & 12.49\sym{**} & -44.00\\
\addlinespace
 & (3.98) & (47.06) & (3.98) & (49.35)\\
\addlinespace
\hspace*{6mm}College x \PARPMTEN &  & 0.72 &  & 0.89\\
\addlinespace
 &  & (0.59) &  & (0.62)\\
\addlinespace
\hspace*{6mm}College x extreme heat &  & 4.15 &  & 1.96\\
\addlinespace
 &  & (2.60) &  & (1.51)\\
\addlinespace
\hspace*{6mm}College x extreme cold &  & -7.54$^\dagger$ &  & -3.65\sym{*}\\
\addlinespace
 &  & (4.17) &  & (1.67)\\
\addlinespace
\addlinespace
\multicolumn{5}{l}{\hspace*{0mm}Control variables}\\
\addlinespace
\hspace*{6mm}Male & 111.10\sym{**} & 111.10\sym{**} & 111.00\sym{**} & 111.00\sym{**}\\
\addlinespace
 & (3.48) & (3.48) & (3.47) & (3.47)\\
\addlinespace
\hspace*{6mm}Mother’s age & 39.56\sym{**} & 39.28\sym{**} & 39.52\sym{**} & 39.27\sym{**}\\
\addlinespace
 & (4.26) & (4.26) & (4.26) & (4.26)\\
\addlinespace
\hspace*{6mm}Mother’s age$^2$ & -0.60\sym{**} & -0.59\sym{**} & -0.60\sym{**} & -0.59\sym{**}\\
\addlinespace
 & (0.07) & (0.07) & (0.07) & (0.07)\\
\addlinespace
\hspace*{0mm}Intercept & 2,193.00\sym{**} & 2,217.00\sym{**} & 2,131.00\sym{**} & 2,159.00\sym{**}\\
\addlinespace
 & (170.90) & (171.90) & (171.90) & (173.10)\\
\addlinespace
\midrule
Observations & 49,973 & 49,973 & 49,973 & 49,973\\
R$^2$	& 0.037 & 0.038 & 0.038 & 0.038\\
\bottomrule
\addlinespace[0.5em]
\multicolumn{5}{p{0.95\textwidth}}{\parbox[t]{0.95\textwidth}{\TABNOTESMAINTABTHREEFULLTERM}}\\
\end{tabular}
\end{adjustbox}
\end{table}

\subsection{Low Birth Weight, Preterm, and Small for Gestational Age\label{sec:appbinary}}

In the text we focus on a continuous measure of birth weight. But in some cases studies focus on related binary variables:  low birth weight (birth weight  under 2500 grams), preterm (gestational age less than 37 weeks), and small for gestational age (1 is defined as those with birth weight under 10 \% in sex-and gestational-age specific distributions). The three binary outcome division strategies could be interpreted geometrically. With gestational-age along the x-axis and birth-weight along the y-axis, low birth weight cuts the data horizontally, preterm cuts the data vertically, and small-for-gestational-age cuts the data diagonally. While each strategy generates valuable outcomes of interest, these data-reduction strategies eliminate much of the distributional variations in the data and preclude quantile analysis. 

In this section, to facilitate comparison of our results to papers that focus on these binary outcomes, we present regression results for low birth weight, preterm and small for gestational age in Tables \ref{tab:mainthreebimargin} and \ref{tab:mainthreebiodds}. Tables \ref{tab:mainthreebimargin} and \ref{tab:mainthreebiodds} differ from Table \ref{tab:mainthree} by replacing least-squares with logistic regressions and replacing birth weight with binary outcome variables, but the same right-side specifications are used. Table \ref{tab:mainthreebimargin} presents results as marginal effects, and Table \ref{tab:mainthreebiodds} presents results as odds-ratios.\footnote{From the summary statistics in Table \ref{tab:mainone}, 6\% of the analytical sample have low birth weight (7\% for high school educated, and 4\% for college educated), 7\% of the analytical sample are borne preterm (8\% for high school educated, and 5\% for college educated), and 9\% are small for gestational age (10\% for high school educated, and 8\% for college educated).} 

The results from Tables \ref{tab:mainthreebimargin} and \ref{tab:mainthreebiodds} largely match-up with results from Table \ref{tab:mainthree}. First, greater \PARPMTEN, extreme heat, as well as extreme cold exposure are all associated with increased probabilities for low birth weight, preterm, and small for gestational age. \PARPMTEN and and extreme heat have the strongest associations with preterm, extreme cold has the strongest association with low birth weight, and the environmental measures are positively but not significantly associated with higher chance for small for gestational age. Second, maternal college education is significantly inversely associated with all three binary outcomes: its strongest association is with low birth weight and its weakest association is with small for gestational age. Third, we find that mothers' college education status is associated with reduced vulnerability to ambient air pollution and extreme heat exposures. Results are similarly for low birth weight and preterm but weaker for small for gestational age. Associations are greater for the extreme heat exposure interaction and are weakly significant for the \PARPMTEN exposure interaction. Interesting, we also find a protective effect of maternal college education for the association between extreme cold and preterm. 

\begin{table}[htbp]
\centering
\captionsetup{width=1.0\textwidth}
\caption{\hspace*{0mm}\TITLETABMAINTHREEBIMARGIN}
\begin{adjustbox}{max width=1.0\textwidth}
\begin{tabular}{m{5.7cm}*{6}{>{\centering\arraybackslash}m{2cm}}}
\toprule
& \multicolumn{6}{c}{Different binary birth-related outcomes}\\
\cmidrule(l{5pt}r{5pt}){2-7}
& \multicolumn{2}{c}{Low birth weight} & \multicolumn{2}{c}{Preterm} & \multicolumn{2}{c}{Small for gestational age}\\
\cmidrule(l{5pt}r{5pt}){2-3} \cmidrule(l{5pt}r{5pt}){4-5} \cmidrule(l{5pt}r{5pt}){6-7}  
Variable & (1) & (2) & (3) & (4) & (5) & (6) \\
\midrule
\addlinespace
\multicolumn{7}{l}{\hspace*{0mm}Environmental exposure variables}\\
\addlinespace
\hspace*{6mm}\PARPMTEN & 0.0066\sym{**} & 0.0067\sym{**} & 0.0080\sym{**} & 0.0082\sym{**} & 0.0010 & 0.0011\\
\addlinespace
 & (0.0011) & (0.0011) & (0.0012) & (0.0012) & (0.0010) & (0.0010)\\
\addlinespace
\hspace*{6mm}Extreme heat & 0.0041 & 0.0053 & 0.0155\sym{**} & 0.0168\sym{**} & 0.0064 & 0.0071\\
\addlinespace
 & (0.0048) & (0.0048) & (0.0057) & (0.0057) & (0.0044) & (0.0044)\\
\addlinespace
\hspace*{6mm}Extreme cold & 0.0156\sym{**} & 0.0157\sym{**} & 0.0127\sym{*} & 0.0145\sym{**} & 0.0094\sym{*} & 0.0087$^\dagger$\\
\addlinespace
 & (0.0049) & (0.0050) & (0.0053) & (0.0053) & (0.0045) & (0.0046)\\
\addlinespace
\addlinespace
\multicolumn{7}{l}{\hspace*{0mm}Education and environmental exposure interactions}\\
\addlinespace
\hspace*{6mm}College educated & -0.0301\sym{**} & 0.0272 & -0.0251\sym{**} & 0.0419 & -0.0195\sym{**} & 0.0012\\
\addlinespace
 & (0.0025) & (0.0301) & (0.0026) & (0.0301) & (0.0029) & (0.0346)\\
\addlinespace
\hspace*{6mm}College x \PARPMTEN &  & -0.0007$^\dagger$ &  & -0.0006 &  & -0.0003\\
\addlinespace
 &  & (0.0004) &  & (0.0004) &  & (0.0004)\\
\addlinespace
\hspace*{6mm}College x extreme heat &  & -0.0043\sym{**} &  & -0.0045\sym{**} &  & -0.0019\\
\addlinespace
 &  & (0.0016) &  & (0.0016) &  & (0.0019)\\
\addlinespace
\hspace*{6mm}College x extreme cold &  & -0.0005 &  & -0.0079\sym{**} &  & 0.0033\\
\addlinespace
 &  & (0.0028) &  & (0.0029) &  & (0.0031)\\
\addlinespace
\addlinespace
\multicolumn{7}{l}{\hspace*{0mm}Control variables}\\
\addlinespace
\hspace*{6mm}Male & -0.0125\sym{**} & -0.0125\sym{**} & 0.0060\sym{**} & 0.0060\sym{**} & 0.0025 & 0.0024\\
\addlinespace
 & (0.0019) & (0.0019) & (0.0021) & (0.0021) & (0.0025) & (0.0025)\\
\addlinespace
\hspace*{6mm}Mother’s age & -0.0141\sym{**} & -0.0139\sym{**} & -0.0144\sym{**} & -0.0143\sym{**} & -0.0108\sym{**} & -0.0107\sym{**}\\
\addlinespace
 & (0.0019) & (0.0019) & (0.0021) & (0.0021) & (0.0028) & (0.0028)\\
\addlinespace
\hspace*{6mm}Mother’s age$^2$ & 0.0002\sym{**} & 0.0002\sym{**} & 0.0003\sym{**} & 0.0003\sym{**} & 0.0002\sym{**} & 0.0002\sym{**}\\
\addlinespace
 & (0.0000) & (0.0000) & (0.0000) & (0.0000) & (0.0000) & (0.0000)\\
\addlinespace
\midrule
Observations & 53,879 & 53,879 & 53,879 & 53,879 & 53,879 & 53,879\\
\bottomrule
\addlinespace[0.5em]
\multicolumn{7}{p{1.23\textwidth}}{\parbox[t]{1.23\textwidth}{\TABNOTESMAINTABTHREEBIMARGIN}}\\
\end{tabular}
\end{adjustbox}
\end{table}

\begin{table}[htbp]
\centering
\captionsetup{width=1.0\textwidth}
\caption{\hspace*{0mm}\TITLETABMAINTHREEBIODDS}
\begin{adjustbox}{max width=1.0\textwidth}
\begin{tabular}{m{5.7cm}*{6}{>{\centering\arraybackslash}m{2cm}}}
\toprule
& \multicolumn{6}{c}{Different binary birth-related outcomes}\\
\cmidrule(l{5pt}r{5pt}){2-7}
& \multicolumn{2}{c}{Low birth weight} & \multicolumn{2}{c}{Preterm} & \multicolumn{2}{c}{Small for gestational age}\\
\cmidrule(l{5pt}r{5pt}){2-3} \cmidrule(l{5pt}r{5pt}){4-5} \cmidrule(l{5pt}r{5pt}){6-7}  
Variable & (1) & (2) & (3) & (4) & (5) & (6) \\
\midrule
\addlinespace
\multicolumn{7}{l}{\hspace*{0mm}Environmental exposure variables}\\
\addlinespace
\hspace*{6mm}\PARPMTEN & 1.137\sym{**} & 1.141\sym{**} & 1.143\sym{**} & 1.146\sym{**} & 1.012 & 1.013\\
\addlinespace
 & (0.001) & (0.001) & (0.001) & (0.001) & (0.001) & (0.001)\\
\addlinespace
\hspace*{6mm}Extreme heat & 1.084 & 1.109 & 1.293\sym{**} & 1.321\sym{**} & 1.081 & 1.090\\
\addlinespace
 & (0.102) & (0.105) & (0.122) & (0.125) & (0.058) & (0.059)\\
\addlinespace
\hspace*{6mm}Extreme cold & 1.357\sym{**} & 1.360\sym{**} & 1.235\sym{*} & 1.272\sym{**} & 1.121\sym{*} & 1.111$^\dagger$\\
\addlinespace
 & (0.132) & (0.132) & (0.108) & (0.111) & (0.061) & (0.062)\\
\addlinespace
\addlinespace
\multicolumn{7}{l}{\hspace*{0mm}Education and environmental exposure interactions}\\
\addlinespace
\hspace*{6mm}College educated & 0.554\sym{**} & 1.704 & 0.660\sym{**} & 2.000 & 0.790\sym{**} & 1.014\\
\addlinespace
 & (0.027) & (1.004) & (0.028) & (0.997) & (0.028) & (0.425)\\
\addlinespace
\hspace*{6mm}College x \PARPMTEN &  & 0.987$^\dagger$ &  & 0.990 &  & 0.996\\
\addlinespace
 &  & (0.007) &  & (0.006) &  & (0.005)\\
\addlinespace
\hspace*{6mm}College x extreme heat &  & 0.920\sym{**} &  & 0.928\sym{**} &  & 0.978\\
\addlinespace
 &  & (0.029) &  & (0.025) &  & (0.023)\\
\addlinespace
\hspace*{6mm}College x extreme cold &  & 0.990 &  & 0.878\sym{**} &  & 1.040\\
\addlinespace
 &  & (0.054) &  & (0.042) &  & (0.039)\\
\addlinespace
\addlinespace
\multicolumn{7}{l}{\hspace*{0mm}Control variables}\\
\addlinespace
\hspace*{6mm}Male & 0.782\sym{**} & 0.782\sym{**} & 1.104\sym{**} & 1.105\sym{**} & 1.030 & 1.030\\
\addlinespace
 & (0.029) & (0.029) & (0.038) & (0.038) & (0.031) & (0.031)\\
\addlinespace
\hspace*{6mm}Mother’s age & 0.759\sym{**} & 0.761\sym{**} & 0.788\sym{**} & 0.789\sym{**} & 0.877\sym{**} & 0.878\sym{**}\\
\addlinespace
 & (0.028) & (0.028) & (0.028) & (0.028) & (0.030) & (0.030)\\
\addlinespace
\hspace*{6mm}Mother’s age$^2$ & 1.005\sym{**} & 1.005\sym{**} & 1.004\sym{**} & 1.004\sym{**} & 1.002\sym{**} & 1.002\sym{**}\\
\addlinespace
 & (0.001) & (0.001) & (0.001) & (0.001) & (0.001) & (0.001)\\
\addlinespace
\midrule
Observations & 53,879 & 53,879 & 53,879 & 53,879 & 53,879 & 53,879\\
\bottomrule
\addlinespace[0.5em]
\multicolumn{7}{p{1.23\textwidth}}{\parbox[t]{1.23\textwidth}{\TABNOTESMAINTABTHREEBIODDS}}\\
\end{tabular}
\end{adjustbox}
\end{table}

\pagebreak 

\renewcommand{\thefigure}{B.\arabic{figure}}
\setcounter{figure}{0}
\renewcommand{\thetable}{B.\arabic{table}}
\setcounter{table}{0}
\renewcommand{\theequation}{B.\arabic{equation}}
\setcounter{equation}{0}
\renewcommand{\thefootnote}{B.\arabic{footnote}}
\setcounter{footnote}{0}

\section{Acquiring Environmental Data\label{sec:appdata}}

A wide array of environmental data is available at detailed geographical units across the globe from the the European Centre for Medium-Range Weather Forecasts (ECMWF)
at the \href{https://cds.climate.copernicus.eu/}{Copernicus Climate Change Service}. 
Copernicus provides data access via a free and publicly accessible \href{https://cds.climate.copernicus.eu/api-how-to}{API service}. In this appendix section, we describe how we obtained key environmental data used in this paper using Copernicus. 

\subsection{Data Retrieval}

Copernicus offers a range of data in different formats with similar data request structures. In particular, temperature as well as other environmental data based on observations from across the globe (with reanalysis) are available from the \href{https://cds.climate.copernicus.eu/cdsapp#!/dataset/reanalysis-era5-pressure-levels?tab=overview}{ERA5 Pressure Level} as well as the \href{https://cds.climate.copernicus.eu/cdsapp#!/dataset/reanalysis-era5-single-levels?tab=overview}{ERA5 Single Level} datasets.

\subsubsection{Single Data Retrieval Request}

To acquire Chinese data for the particular period in which our birth outcome data are available, we need to specify the appropriate time ranges as well as the geographical coordinates. We retrieve hourly data from every day between the year 2007 and 2012 by specifying the appropriate \emph{year}, \emph{month}, \emph{day}, and \emph{time} parameters. We specify our data acquisition geographical area as to the south-east of latitude and longitude coordinates (in decimal degrees) $\left(23.50, 113.00\right)$ and to the north-west of coordinates $\left(22.25, 114.50\right)$, which covers the broad geographical area that is relevant for our paper. Our specification for the \emph{area} parameter is therefore $[23.50, 113.00, 22.25, 114.5]$. 

Given this information and after registering with Copernicus to obtain an  user-specific url and passkey, Source Code \ref{code:codeone} provides a API call to acquire temperature data from Copernicus in \emph{grib} format.

\definecolor{dhscodebg}{rgb}{0.95,0.95,0.95}

\begin{listing}[H]
\caption{Single Data Retrieval Call\label{code:codeone}}
\begin{minted}
[
frame=lines,
xleftmargin=2mm,
framesep=2mm,
baselinestretch=1.2,
bgcolor=dhscodebg,
fontsize=\scriptsize,
linenos
]
{python3}
# Library
import cdsapi
import urllib.request

# download folder
spt_root = "C:/data/"
spn_dl_test_grib = spt_root + "test_china_temp.grib"
# request
c = cdsapi.Client()
res = c.retrieve("reanalysis-era5-pressure-levels",
         {
             'product_type': 'reanalysis',
             'variable': 'temperature',
             'pressure_level': '1000',
             'year': [ '2007', '2008','2009', '2010', '2011', '2012' ]
             'month': [ '01','02','03','04','05','06', '07','08','09','10','11','12'],
             'day': [
                '01','02','03','04','05','06','07','08','09','10','11','12',
                '13','14','15','16','17','18','19','20','21','22','23','24',
                '25','26','27','28','29','30','31'
             ],
             'time': [
                '00:00', '01:00', '02:00', '03:00', '04:00', '05:00', '06:00', '07:00', '08:00',
                '09:00', '10:00', '11:00', '12:00', '13:00', '14:00', '15:00', '16:00', '17:00',
                '18:00', '19:00', '20:00', '21:00', '22:00', '23:00'
             ],
             'area': [23.50, 113.00, 22.25, 114.5],
             'grid': [1.25, 0.25],
             "format": "grib"
         },
         spn_dl_test_grib
         )

\end{minted}
\end{listing}

\subsubsection{Subdivided Data Retrieval Request}

A challenge to taking full advantage of the data is that, given the fine geographical and time units, the resulting data files can become very large. A single call to acquire all relevant data as shown in the example above can only be implemented on a server with access to Terabytes of storage space. 

To deal with this challenge, we divide our API calls into smaller components. We make multiple requests of data at shorter time intervals. Each time we aggregate and process the relevant data before downloading the next set of data. Given the computational resources at our disposal, we download the data at six months intervals, as shown in Source Code \ref{code:codetwo}. Given the computing resources available to the researcher, the time intervals can be further shortened to circumvent computational challenges from using the data.

\begin{listing}[H]
\caption{Sub-Period Data Retrieval Call\label{code:codetwo}}
\begin{minted}
[
frame=lines,
xleftmargin=2mm,
framesep=2mm,
baselinestretch=1.2,
bgcolor=dhscodebg,
fontsize=\scriptsize,
linenos
]
{python3}
# date lists
ar_years = 2001:2019
ar_months_g1 = ['01','02','03','04','05','06']
ar_months_g2 = ['07','08','09','10','11','12']

# Loop over time periods
for it_yr in ar_years:
    for it_mth_group in [1, 2]:
        if it_mth_group == 1:
            ar_months = ar_months_g1
        if it_mth_group == 2:
            ar_months = ar_months_g2

        c = cdsapi.Client()
        res = c.retrieve(
            'reanalysis-era5-pressure-levels',
            {
                'product_type': 'reanalysis',
                'variable': 'temperature',
                'pressure_level': '1000',
                'year': [it_yr],
                'month': ar_months,
                 'day': [
                    '01','02','03','04','05','06','07','08','09','10','11','12',
                    '13','14','15','16','17','18','19','20','21','22','23','24',
                    '25','26','27','28','29','30','31'
                 ],
                'time': [
                    '00:00', '01:00', '02:00', '03:00', '04:00', '05:00', '06:00', '07:00', '08:00',
                    '09:00', '10:00', '11:00', '12:00', '13:00', '14:00', '15:00', '16:00', '17:00',
                    '18:00', '19:00', '20:00', '21:00', '22:00', '23:00'
                ],
                'area': [23.50, 113.00, 22.25, 114.5],
                'grid': [0.25, 0.25],
                'format': 'grib'
            },
            "china_temp.grib")
\end{minted}
\end{listing}

\subsection{Data Processing}

The data we download is at finer detail than required by the statistical analysis. For each sub-period of data downloaded, we process the data using a variety of tools. Data in the \emph{grib} format is processed using the \href{https://xarray.pydata.org/en/stable/}{xarray} package as shown in Source Code \ref{code:codethree}. Data in \emph{netCDF} format is processed using the \href{https://pypi.org/project/netCDF4/}{netCDF4} package as shown in Source Code \ref{code:codefour}. We store the resulting aggregate data as csv files and combine that with the rest of our child birth outcome data to conduct relevant statistical analysis.

\begin{listing}[H]
\caption{Grib Data Processing with xarray\label{code:codethree}}
\begin{minted}
[
frame=lines,
xleftmargin=2mm,
framesep=2mm,
baselinestretch=1.2,
bgcolor=dhscodebg,
fontsize=\scriptsize,
linenos
]
{python3}
# Load Packages
import pandas as pd
import xarray as xr

# Process grid data
snm_data_grib, snm_data_csv = "data.grib", "data.csv"
dsxr = xr.load_dataset(snm_data_grib, engine='cfgrib')
pd.concat([dsxr['u10'].to_series(), dsxr['v10'].to_series(),
           dsxr['d2m'].to_series(), dsxr['t2m'].to_series(),
           dsxr['msl'].to_series(), dsxr['sp'].to_series()],
          axis=1).to_csv(snm_data_csv, index=True)
\end{minted}
\end{listing}

\begin{listing}[H]
\caption{netcdf Data Processing with netCDF4\label{code:codefour}}
\begin{minted}
[
frame=lines,
xleftmargin=2mm,
framesep=2mm,
baselinestretch=1.2,
bgcolor=dhscodebg,
fontsize=\scriptsize,
linenos
]
{python3}
# Load Packages
import pandas as pd
from netCDF4 import Dataset, date2num, num2date

# Process netCDF data
snm_data_nc, snm_data_csv = "data.nc", "data.csv"
ds_src = Dataset(snm_data_nc)
var_tp = ds_src.variables['tp']

# Get the three dimensions, time, lat, and long
time_dim, lat_dim, lon_dim = var_tp.get_dims()
time_var = ds_src.variables[time_dim.name]
times = num2date(time_var[:], time_var.units)

# The flattening at the end converts variables to single column
latitudes = ds_src.variables[lat_dim.name][:]
longitudes = ds_src.variables[lon_dim.name][:]

# Convert to dataframe
[mt_times, mt_lat, mt_long] = np.meshgrid(times, latitudes, longitudes, indexing='ij')
ar_times = np.ravel(mt_times)
ar_lat = np.ravel(mt_lat)
ar_long = np.ravel(mt_long)
df = pd.DataFrame({'time': [t.isoformat() for t in ar_times],
                   'latitude': ar_lat, 'longitude': ar_long, 'tp': var_tp[:].flatten()})

# Get date and hour
df['date'] = pd.to_datetime(df['time']).dt.date
df['hour'] = pd.to_datetime(df['time']).dt.hour

# sort and group, and summ
sr_day_sum = df.groupby(['latitude','longitude','date'])['tp'].sum()
df_day_sum = sr_day_sum.reset_index()

# convert to csv
df_day_sum.to_csv(snm_data_csv, index=False)
\end{minted}
\end{listing}

\end{document}